\documentclass[aps,prb,amsmath,amssymb,footinbib,longbibliography,superscriptaddress,twocolumn]{revtex4-2}

\usepackage{t1enc}
\usepackage[utf8]{inputenc}
\usepackage{dsfont}
\usepackage[normalem]{ulem} 
\usepackage{nicematrix}
\usepackage{tikz}
\usepackage{cancel}
\usepackage{mathtools}
\usepackage{booktabs}
\usepackage{multirow}

\usepackage{graphicx,bm,color} 

\usepackage[unicode=true, colorlinks=true, citecolor={blue!80!black}, urlcolor={blue!50!black}, linkcolor = {blue!80!black}]{hyperref}
\usepackage{physics}
\usepackage{bbm}

\usepackage{soul}
\usepackage[english]{babel}

\begin{document}
\title{Time-optimal quantum gates with bang-bang control in multilevel systems}
\newcommand{\affA}{\affiliation{Grupo de Circuitos Cuánticos Bariloche, Div. Dispositivos y Sensores, Centro Atómico Bariloche-CNEA, 8400 San Carlos de Bariloche, Argentina.}}
\newcommand{\affB}{\affiliation{Centro At\'omico Bariloche and Instituto Balseiro, 8400 San Carlos de Bariloche, R\'io Negro, Argentina.}}
\newcommand{\affC}{\affiliation{Instituto de Nanociencia y Nanotecnolog\'{\i}a (INN), CONICET-CNEA, Argentina.}}

\author{Valentín Reparaz}
\affA
\author{Santiago Ferreyra}
\affB
\author{María José Sánchez}
\affB
\affC
\author{Daniel Domínguez}
\affB
\author{Leandro Tosi}
\email[Corresponding author: ]{leandro.tosi@ib.edu.ar}
\affA
\affB
\date{\today}

\begin{abstract}
We determine the optimal quantum manipulation protocols for implementing high-fidelity, fast single-qubit gates. We demonstrate that the time-optimal pulse sequence is a ``bang-bang'' sequence: discrete pulses of either positive or negative maximum amplitude or zero. The non-adiabatic bang-bang pulse sequence minimizes gate duration providing a speedup for low-frequency architectures. We first derive the protocol for a transversally driven two-level system, extracting exact analytic expressions for minimum gate times. We then extend this framework to general multilevel architectures, identifying conditions that enable the coherent suppression of leakage errors. Using the fluxonium circuit as a representative case study, we optimize $X/2$ and $Y/2$ gates through a combination of discrete bang sequences and continuous waveform smoothing. This approach preserves near-optimal execution speeds while mitigating transitions outside the computational subspace. Open-system simulations demonstrate that these sequences outperform commensurate and resonant pulse schemes across different fluxonium regimes, achieving low-error manipulation significantly faster than standard resonant control, even in the presence of $1/f$ flux noise and dissipation.
\end{abstract}
\maketitle

\section{\label{sec:intro}Introduction}
Historically, coherent manipulation protocols for quantum systems have borrowed heavily from the toolkit of nuclear magnetic resonance (NMR) \cite{vandersypen2004nmr}. The direct mapping of a two-level system (TLS) to a spin-1/2 particle naturally reinforced this trend, establishing foundational concepts such as the Rabi model alongside ubiquitous control sequences like $\pi$-pulses, $\pi$/2-pulses, Ramsey interferometry, and Hahn spin-echoes \cite{hahn1950spin, ramsey1950molecular, slichter2013principles}. Different physical architectures, notably superconducting quantum circuits \cite{blais2021circuit, kjaergaard2020superconducting}, have subsequently developed their own dictionaries to translate this established NMR language into specific experimental control knobs \cite{gu2017microwave, blais2004cavity}. However, in the ongoing quest to implement high-fidelity, fast quantum gates, we return to first principles to address a fundamental question: What is the strictly optimal manipulation protocol? 

The coherent control of quantum bits is conventionally achieved via resonant microwave drives. In the weak-drive regime, the rotating-wave approximation (RWA) holds, and the duration of a resonant quantum gate scales inversely with the drive amplitude \cite{nielsen2000quantum, vandersypen2004nmr}. However, faster control requires stronger drives, which invalidate the RWA by activating counter-rotating terms \cite{krantz2019quantum, lizuain2008vibrational} and induce leakage errors \cite{motzoi2009simple, chen2016measuring} in multilevel architectures. The resulting speed limit is particularly detrimental for low-frequency qubits \cite{manucharyan2009fluxonium, somoroff2023millisecond, nguyen2019high} which are highly attractive due to their extended coherence times, but typically suffer from gate durations that are too long for practical applications.

In the strong-drive regime, Landau-Zener-Stückelberg (LZS) \cite{zhang2021universal, campbell2020universal, caceres2023fast, ferreyra2026optimizing} and commensurate-pulse protocols \cite{rower2024suppressing} have demonstrated a considerable reduction of gate time. These non-adiabatic approaches are particularly suited for quantum systems with large anharmonicity. However, identifying the absolute quantum speed limit for these gate protocols remains an open challenge.

To establish these fundamental bounds, we must depart from heuristic pulse design and turn to Quantum Optimal Control (QOC). While QOC is extensively employed to engineer robust, high-fidelity pulse shapes \cite{brif2010control, koch2016controlling, muller2022one, kallush2014quantum}, our objective is distinctly to determine the absolute quantum speed limits of these gates \cite{khaneja2001time}. We approach this by applying Optimal Control Theory (OCT) and Pontryagin's Principle \cite{vinter2021optimal, boscain2021introduction} to drastically constrain the optimization problem. Rather than searching through an unconstrained continuous waveform space, Pontryagin's Principle establishes that for a control-affine system (where the control appears linearly in the Hamiltonian) under bounded control, the time-optimal driving field is not sinusoidal, but rather a ``bang-bang'' sequence: discrete pulses of either positive or negative maximum amplitude or zero. While time-optimal state transfer of a TLS is well studied \cite{boscain2006time, avinadav2014time, evangelakos2023minimum, hirose2018time}, the optimization for full gate fidelity has been developed only recently. Previously, gate optimization was restricted to $\pi$ rotations \cite{lin2025time}, which together with $Z$ rotations do not generate universal single-qubit gates. 

Moreover, because these approaches rely on strong broadband pulses, they break the rotating-frame formalism required to synthesize instantaneous virtual $Z$ gates in software \cite{mckay2017efficient}. It is thus crucial to determine whether the non-adiabatic gate speedups outweigh the loss of this highly efficient control technique.

The challenge we address is to establish a control paradigm capable of synthesizing $\pi/2$ transverse gates for realistic, noisy, potentially multilevel qubit architectures. To this end we develop a theoretical and numerical framework for the synthesis of time-optimal, non-adiabatic gates based on bang-bang pulse sequences. We formulate a comprehensive protocol for $\pi/2$ gates and implement it first for a TLS and then for a fluxonium, a realistic multilevel superconducting circuit architecture. Our methodology applies to any quantum system with sufficiently large anharmonicity. We demonstrate that fast, high-fidelity $\pi/2$ rotations can be achieved by utilizing a bang-idle-bang and three-bang pulse sequence for $Y/2$ and $X/2$ gates, respectively. We analyze the resulting leakage landscape and find conditions for coherent suppression of leakage, which mitigates this type of error and allows for high fidelity. Furthermore, we show that applying minimal continuous smoothing to these discrete bang-bang sequences preserves near-optimal gate speeds while improving experimental feasibility and reducing leakage.

\begin{figure*}[t!]
    \centering
    \includegraphics[width=0.9\textwidth]{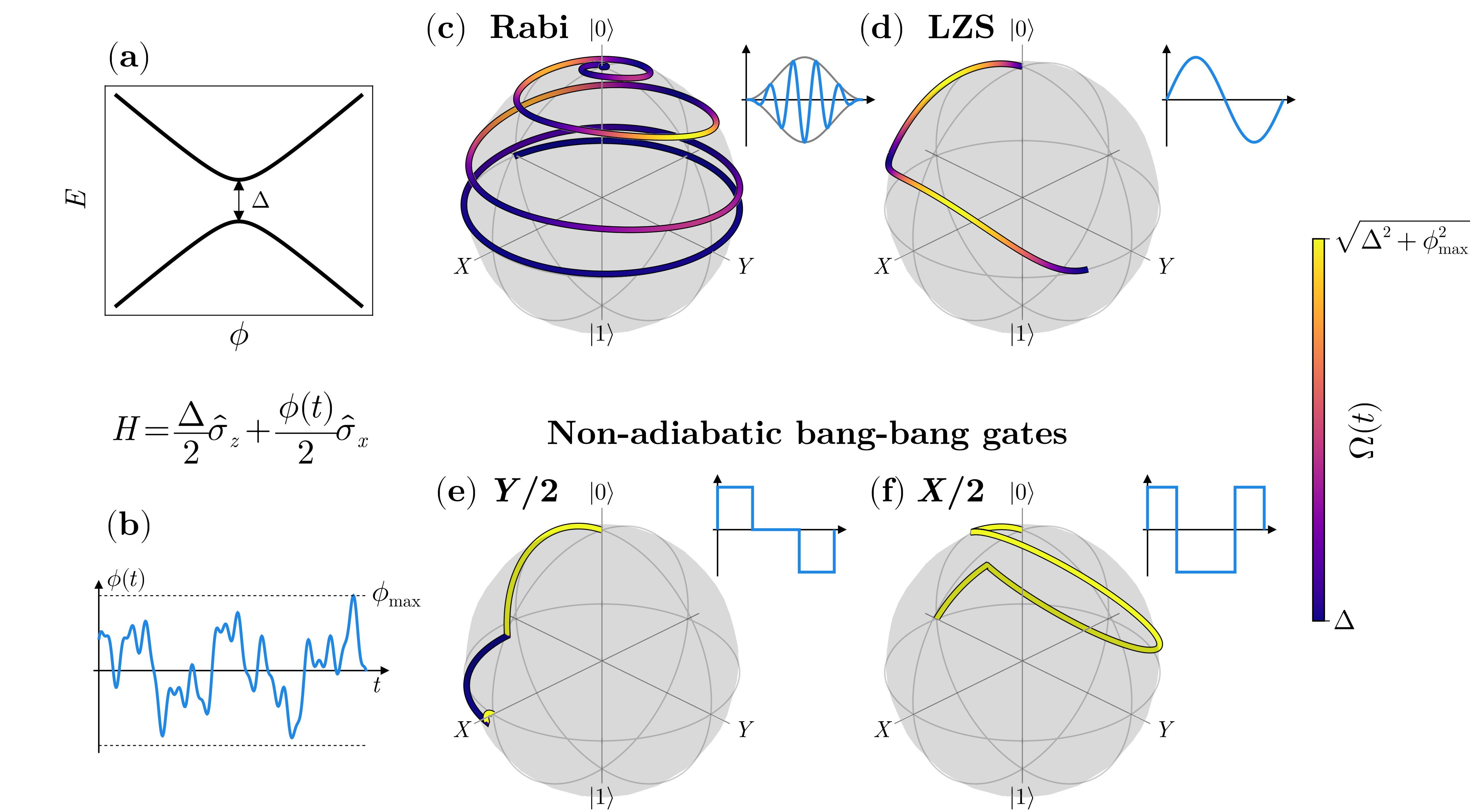}
    \caption{\textbf{Single-qubit gate schemes under bounded transverse control.} (a) Energy spectrum of a transversally driven two-level system with gap $\Delta$. (b) The optimization problem: determining the time-optimal control field $\phi(t)$ bounded by $\pm\phi_{\text{max}}$ to execute a target $\pi/2$ gate. (c-f) Bloch sphere trajectories for the initial state $\ket{0}$ under different control schemes. Insets display the corresponding drive waveforms $\phi(t)$. The trajectory color maps the instantaneous rotation speed $\Omega(t) = \sqrt{\Delta^2 + \phi(t)^2}$ around the axis $\hat{n} \propto (\phi(t), 0, \Delta)$. (c) Resonant (Rabi) gate. In the weak-drive regime ($\phi_{\text{max}} \ll \Delta$), the field must oscillate in resonance with the Larmor precession to accumulate a meaningful rotation. (d) Landau-Zener-Stückelberg (LZS) gate. For low-frequency qubits, resonant gates become impractically slow. By driving harder, gates become possible even when not in resonance. Pre- and post idling is required to obtain the correct final phase. (e-f) Time-optimal, non-adiabatic bang-bang gates emerging in the strong-drive regime. These are the extreme limit of LZS control, saturating the drive to rotate the state as fast as possible for as much time as possible. (e) The optimal $Y/2$ gate consists of two saturated pulse (yellow curves) separated by an idling period (blue arc). (f) The time-optimal $X/2$ gate requires a three-bang sequence to correct for the small $Z$-drift.}
    \label{fig:fig1}
\end{figure*}

The manuscript is organized as follows: in Sec. II, we describe the dynamics of strongly driven two-level systems and use Pontryagin's Principle to optimize for minimum-time $\pi/2$ gates. In Sec. III, we introduce the multilevel fluxonium circuit and detail the mechanism of coherent suppression of leakage. In Sec. IV, we demonstrate the synthesis of high-fidelity fluxonium gates incorporating pulse smoothing, benchmarking their performance across different fluxonium regimes.

\section{Time-Optimal Control of Two-Level Systems}

\subsection{Strongly driven TLS}

Single-qubit gates are modeled via the Hamiltonian of a transversally driven TLS,
\begin{equation}
    \hat{H}(t) = \frac{\Delta}{2} \hat{\sigma}_z + \frac{\phi(t)}{2} \hat{\sigma}_x,
    \label{eq:H_TLS}
\end{equation}
where $\Delta$ is the qubit angular frequency, $\phi(t)$ an external tunable parameter (see Fig. \ref{fig:fig1}(a)), and we take $\hbar=1$. The effect of the control field $\phi(t)$ is best understood geometrically by tracking a state vector on the Bloch sphere. In the absence of a drive, the Hamiltonian is diagonal, and an arbitrary state vector precesses around the $Z$-axis with a Larmor period $\tau_L = 2\pi/\Delta$. When the drive $\phi(t)$ is applied, the state rotates around the instantaneous axis $\hat{n}(t)$:
\begin{equation}
    \hat{n}(t) = \frac{1}{\Omega(t)}(\phi(t), 0, \Delta),
\end{equation}
with instantaneous angular speed $\Omega(t) = \sqrt{\Delta^2 + \phi(t)^2}$. As shown in Fig. \ref{fig:fig1}(b), the optimization problem is to identify the control field $\phi(t)$ that executes a desired rotation to an arbitrary state with low error and minimal time. 

Single-qubit control is conventionally implemented using harmonic drives in the weak-drive regime ($\phi_\text{max} \ll \Delta$)  (with $\phi_{\text{max}}$ the drive amplitude) such that the instantaneous axis $\hat{n}(t)$ barely deviates from the $Z$-axis. As shown in Fig. \ref{fig:fig1}(c), to execute a transverse rotation, the drive must oscillate together with the qubit's Larmor precession. This synchronicity allows the qubit to constructively accumulate phase, yielding the standard Rabi $\pi/2$ gate with a duration $t_g \propto \pi/\phi_{\text{max}}$. In high-frequency qubit architectures, the $\phi_{\text{max}} \ll \Delta$ condition is easily satisfied; the large $\Delta$ permits drive amplitudes $\phi$ that yield fast gate times while safely allowing the RWA, which neglects rapidly oscillating counter-rotating terms, to hold true. In multilevel architectures, the driving amplitude is limited by leakage to non-computational levels, which can be partially mitigated using techniques such as DRAG \cite{motzoi2009simple, hyyppa2024reducing}.
To achieve fast gates (decreasing $t_g$) with low-frequency qubits, the drive amplitude must be pushed into the strong-drive regime ($\phi_\text{max} \sim \Delta$). For resonant drives, increasing the amplitude invalidates the RWA, which degrades coherent gate fidelity. While it is not strictly necessary to abandon sinusoidal waveforms, as demonstrated by LZS gate protocols \cite{campbell2020universal, caceres2023fast, ferreyra2026optimizing} that utilize strong single-period sinusoids (see Fig. \ref{fig:fig1}(d)), their continued use is largely a legacy of microwave control. Once the weak-drive limit is lost, the RWA breaks down and the resonance condition becomes irrelevant, there is no fundamental reason to restrict the control field $\phi(t)$ to sinusoidal waveforms. Because these non-adiabatic pulses bypass the rotating frame entirely, we refer to them as lab-frame gates or lab-frame control.

\subsection{Optimal Control Theory}

To bypass the limitations of weak resonant drives, we examine the general optimal control problem illustrated in Fig. \ref{fig:fig1}(b). Given a bounded drive $\phi(t) \in [-\phi_{\text{max}}, \phi_{\text{max}}]$, what is the optimal waveform that executes a target $\pi/2$ gate in the absolute minimum time? By employing Pontryagin's Principle (PP) for time-optimization \cite{vinter2021optimal, lin2025time}, we drastically reduce the solution's space from arbitrary continuous waveforms $\phi(t)$ to a highly constrained family of pulses. The optimal sequence is then selected from this family to simultaneously minimize execution time and maximize fidelity. To quantify this, we define the average gate fidelity between the synthesized unitary $U$ and a target unitary $V$ for a $d$-dimensional system ($d=2$ for a TLS) as
\begin{equation}
    F(U, V) = \frac{\frac{1}{d} |{\rm tr}(U^\dag V)|^2 + 1}{d+1}.
    \label{eq:fidelity_TLS}
\end{equation} 
The application of PP to Eq. \eqref{eq:fidelity_TLS} shows that time-optimal solutions consist of maximum-amplitude ``bangs'' ($\phi(t) = \pm\phi_{\text{max}}$) and zero-amplitude singular arcs, or ``idles'' ($\phi(t) = 0$). Furthermore, we demonstrate in Appendix \ref{ap:pontryagin} that: (i) the only time-optimal protocol featuring an idling segment is the bang-idle-bang sequence ($N=2$ bangs); and (ii) any multi-bang sequence with more than three bangs ($N \ge 3$) is forced to have an identical duration $\tau_m$ for all inner bangs. While (ii) was a known result in literature \cite{lin2025time}, (i) is a novel result from our work.

We further reduce the pulses families by imposing the global condition $F=1$. Achieving a target gate $V=X/2$ ($Y/2$) requires the waveform $\phi(t)$ to have an even (odd) symmetry with respect to its midpoint $t_g/2$, dictating that $N$ must be odd (even). With all these considerations, the number of parameters for the optimization reduces to four, namely: the sign of the first bang, its duration $\tau_1$ (equal to the last bang, $\tau_N=\tau_1$), the total number of bangs $N$, and the duration of the intermediate bangs $\tau_m$. In Fig. \ref{fig:fig1}(e) we show the evolution of the state in the Bloch sphere for a bang-idle-bang sequence that implements a $Y/2$ gate with ideal fidelity. The sequence has $N=2$ bangs, is odd with respect to the midpoint and presents an idle in the middle. In Fig. \ref{fig:fig1}(f) the implementation of a $X/2$ gate is realized with a three-bang ($N=3$) sequence, even with respect to the midpoint.

\subsection{Numerical optimization}

\begin{figure}[t!]
    \centering
\includegraphics[width=0.5\textwidth]{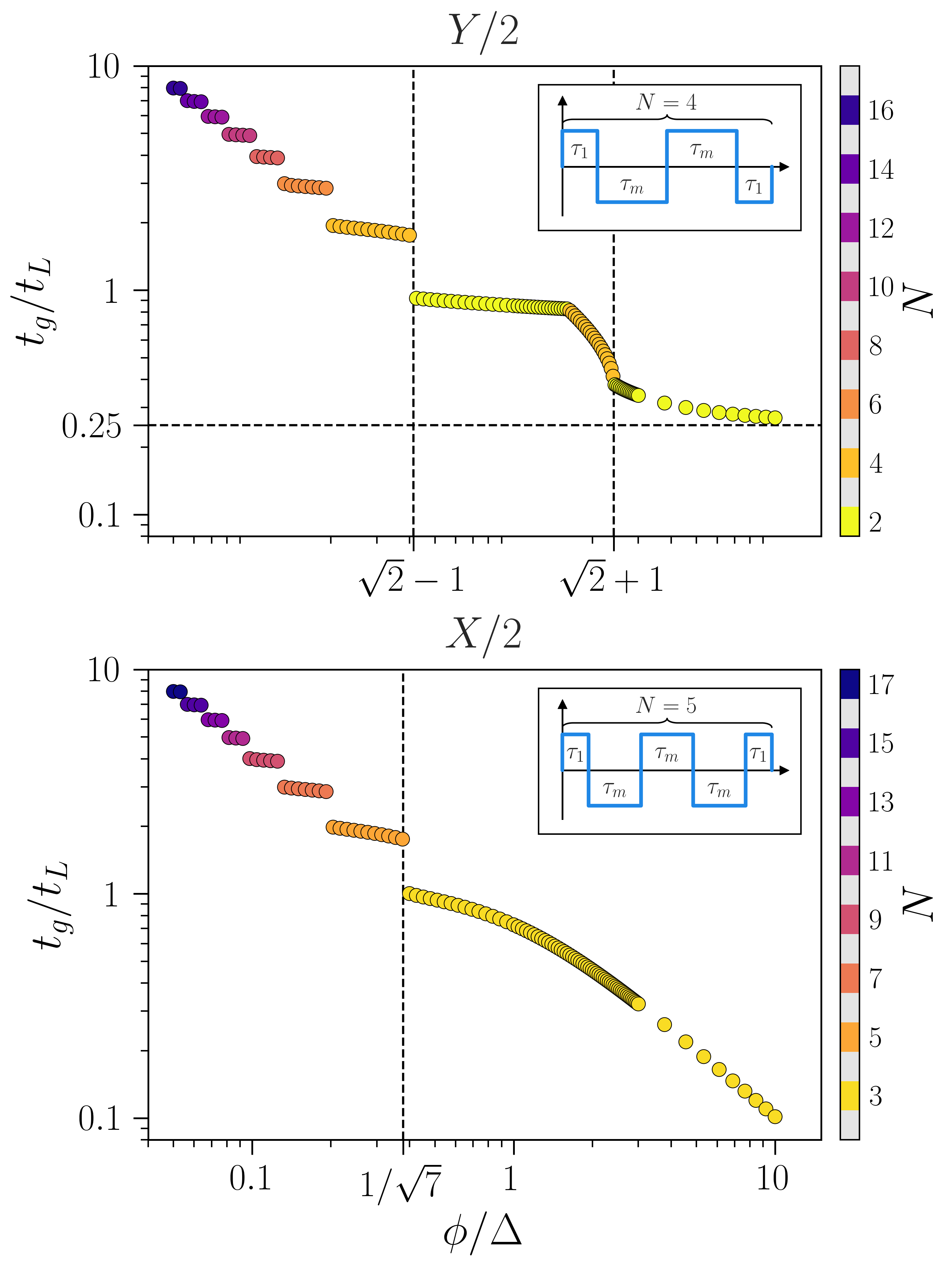}
    \caption{\textbf{Time-optimal gates.} Minimum gate time $t_g$ (normalized by the Larmor period $\tau_L = 2\pi/\Delta$) as a function of drive amplitude $\phi/\Delta$ for $Y/2$ (top) and $X/2$ (bottom) rotations. Point color indicates the optimal number of bangs, $N$. For low amplitudes, $N$ takes on higher values, recovering the resonant gate limit. \textbf{Insets:} Schematics of the optimal bang-bang protocols, illustrated for the specific cases of $N=4$ (top) and $N=5$ (bottom). Sequences alternate between amplitudes $\pm\phi$, parameterized by outer bang durations $\tau_1$ and inner bang durations $\tau_m$. $Y/2$ gates require an even $N$, while $X/2$ gates require an odd $N$. Dashed lines denote analytically derived thresholds in the small-$N$ regime. The points with $N=2$ mark the only case where a finite idling interval is required for the time-optimal solution.}
    \label{fig:fig2}
\end{figure}

Given a value of $\phi_{\text{max}}/\Delta$, which we hereafter denote simply as $\phi/\Delta$, being the only value the drive takes,  we optimize the durations of the outer ($\tau_1$) and middle ($\tau_m$) bangs considering various total bang numbers $N$. These parameters are optimized to satisfy the fidelity constraint $F(U(t_g), V) = 1$ for target gates $V \in \{X/2, Y/2\}$. This search is performed along contours of increasing total gate time $t_g$ until a perfect-fidelity solution is found, ensuring a global optimal time (see optimization details provided in Appendix \ref{ap:optimization}).
The results for the $X/2$ and $Y/2$ target gates are shown in Fig. \ref{fig:fig2}. In both cases, we show the optimal gate time in units of the Larmor period $t_g/\tau_L$ as a function of $\phi/\Delta$. The color indicates the number of bangs in the optimal pulse sequence.

In the weak-drive limit ($\phi/\Delta \ll 1$), the time-optimal solution requires a longer sequence of bangs ($N \gg 1$). The lengths of the outer bangs become negligible, while the inner bang durations converge to $\tau = \pi/\omega_\text{eff}$, with $\omega_\text{eff} \approx \Delta$. The time-optimal sequence recovers resonant control, executing a discretized version of a resonant pulse $\sim$20\% faster than an equivalent sinusoidal pulse of amplitude $\phi$ \cite{lin2025time}.

As $\phi$ increases, the required number of bangs $N$ decreases, together with abrupt reductions in gate time. For the $Y/2$ gate, analytic expressions show that when $\phi/\Delta > \sqrt{2}+1$, the optimal solution is the $N=2$ bang-idle-bang sequence, with a positive initial bang (see App. \ref{ap:gate_analytics}). As illustrated in Fig. \ref{fig:fig1}(e) geometrically, the sequence involves an initial rotation around an axis nearly aligned with $X$, followed by an idle period for phase accumulation, and a final negative bang to reach the target state. Note that unlike simple state transfer, where a single bang-idle sequence would be sufficient, two bangs are necessary here to preserve the symmetry required for full gate fidelity. Gate time is limited by the $Z$-precession during the idle arc, setting the theoretical minimum of $t_g = \tau_L/4$. Another optimal solution with $N=2$ exists for $\phi/\Delta > \sqrt{2}-1$ (see vertical dashed line in Fig. \ref{fig:fig2}) corresponding to an initial bang of negative amplitude. This remains the fastest solution up to a critical ratio of $\phi/\Delta$, beyond which the required idling duration becomes too time-consuming, rendering a sequence with $N=4$ bangs strictly optimal.

Time-optimal sequences for an $X/2$ gate exhibit a different structure. Achieving $F=1$ fast with a single bang is impossible due to the intrinsic $Z$-drift induced by the $\Delta \hat{\sigma}_z$ term in Eq. \eqref{eq:H_TLS}. Since OCT prohibits corrective idling periods (App. \ref{ap:pontryagin}), the control sequence must consist of at least three bangs ($N \ge 3$). Above the analytical threshold $\phi/\Delta > 1/\sqrt{7}$, the $N=3$ sequence is the fastest protocol. As illustrated in Fig. \ref{fig:fig1}(f), the outer positive bangs briefly steer the state, positioning it such that the central negative bang can perform the appropriate rotation around the Bloch sphere and land with the correct phase. Because odd-$N$ optimal protocols do not include idle segments, their execution time has no theoretical lower bound, scaling asymptotically as $t_g \propto \left( \frac{\phi}{\Delta} \right)^{-1}$ when $\phi/\Delta \rightarrow \infty$.

While these sequences set the theoretical gate time limits, infinitely sharp rectangular pulses cannot be synthesized by standard arbitrary waveform generators (AWGs) and require smoothing \cite{hirose2018time}. Furthermore, although these protocols are applicable to native two-level systems \cite{morton2006bang, avinadav2014time}, subjecting multilevel architectures to such extreme drive amplitudes produces leakage \cite{chen2016measuring,motzoi2009simple} and exposes the qubit to first-order $1/f$ flux noise \cite{ithier2005decoherence}. To viably implement these time-optimal non-adiabatic sequences, we must go beyond the TLS approximation and adapt them for a realistic, highly anharmonic multilevel architecture, accounting for experimental factors such as noise, dissipation, and pulse smoothing. We first  turn to the fluxonium circuit to analyze the leakage effects induced by these bang-bang drives.

\section{Fluxonium Circuit and Coherent Suppression of Leakage}

\begin{figure*}[t!]
    \centering
\includegraphics[width=\textwidth]{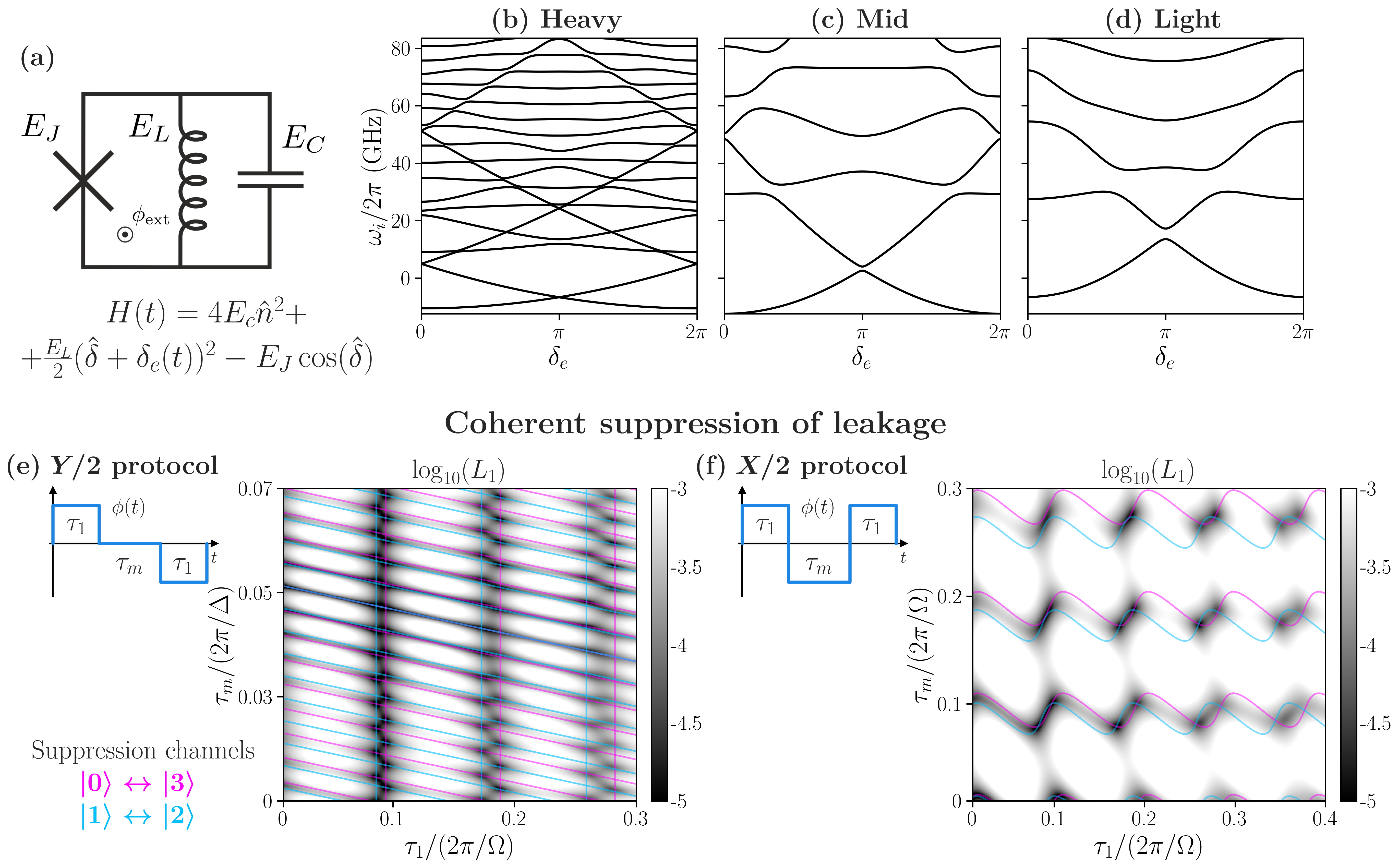}
    \caption{\textbf{Beyond two levels: fluxonium regimes and coherent suppression of leakage.} (a) Circuit diagram and Hamiltonian of the fluxonium qubit, controlled via an external magnetic flux parameterized as $\delta_e = 2\pi\phi_{\text{ext}}/\Phi_0$. (b-d) Energy spectra $\omega_i$ versus external flux for three parameter regimes $(E_C/h, E_L/h, E_J/h)$: (b) Heavy ($0.479, 0.132, 3.395$ GHz) yielding $\Delta/2\pi = 14$ MHz and anharmonicity $\alpha = 214$; (c) Mid ($1.30, 0.59, 5.71$ GHz) yielding $\Delta/2\pi = 222$ MHz and $\alpha = 23$; and (d) Light ($1, 1, 4$ GHz) yielding $\Delta/2\pi = 580$ MHz and $\alpha = 5$. \textbf{Coherent suppression of leakage:} (e-f) State-averaged leakage $L_1$ for the heavy fluxonium under bang-bang (e) $Y/2$ and (f) $X/2$ protocols as a function of the normalized outer ($\tau_1$) and inner ($\tau_m$) bang lengths. The drive amplitude is set to $\phi/\Delta=20$, where $\phi$ represents the effective two-level transverse drive strength. Solid lines denote analytical zeroes for specific leakage channels (obtained in App. \ref{ap:gate_analytics}) dictated by selection rules at the sweet spot: $\ket{0} \leftrightarrow \ket{3}$ (pink) and $\ket{1} \leftrightarrow \ket{2}$ (blue). Intersections of these curves yield total leakage suppression.}
    \label{fig:fig3}
\end{figure*}

\subsection{Fluxonium Hamiltonian}
The fluxonium is a superconducting circuit architecture (see Fig. \ref{fig:fig3}(a)) highly suitable for strong driving due to its large anharmonicity and low transition frequency. Its dynamics is governed by the Hamiltonian
\begin{equation}
    \hat{H}_0 = 4E_C \hat{n}^2 + \frac{E_L}{2} (\hat{\delta} + \delta_e)^2 - E_J \cos\hat{\delta},
    \label{eq:H_fluxonium}
\end{equation}
where the conjugate operators $\hat{n}$ and $\hat{\delta}$ satisfy $[\hat{\delta},\hat{n}]=i$, and represent the charge number and phase degrees of freedom, respectively. The spectrum of the fluxonium and its dynamics are determined by the relative value of the charging energy $E_C$, the inductive energy $E_L$, and the Josephson energy $E_J$. As the control knob we consider the external magnetic flux threading the loop $\phi_\text{ext}$, parameterized as $\delta_e = 2\pi \phi_\text{ext}/\Phi_0$, in units of the flux quantum $\Phi_0$. In Fig. \ref{fig:fig3}(b-d) we present the spectrum of the lowest energy levels as a function of $\delta_e$ for different values of $(E_C,E_L,E_J)$. We investigate three distinct fluxonium parameter regimes well established in the literature which we name Heavy \cite{zhang2021universal}, Mid \cite{rower2024suppressing}, and Light \cite{nguyen2022blueprint}, spanning a wide range of qubit transition frequencies $\Delta/2\pi = 14-580\;\text{MHz}$ and anharmonicities $\alpha = \frac{\omega_2-\omega_1}{\Delta}$ ranging from 214 to 5 (where $\omega_n$ is the energy of the n-th level, $n=0,1,\dots$). The logical qubit is encoded in the lowest two eigenstates at the half-flux quantum ``sweet spot'' ($\delta_e=\pi$), where $\ket{0}$ is the ground state and $\ket{1}$ is the first excited state, separated by the qubit frequency $\Delta=\omega_1-\omega_0$. At this operating point, the circuit is protected against first-order flux noise \cite{manucharyan2009fluxonium}, and its low transition frequency suppresses dielectric relaxation \cite{pop2014coherent}, yielding long coherence times ($T_2$ and $T_1$ respectively) \cite{somoroff2023millisecond, nguyen2019high}.

Coherent manipulation is realized by a fast antenna such that $\delta_e \rightarrow \pi+\delta_e(t)$ leads to the time-dependent Hamiltonian
\begin{equation}
    \hat{H}(t) =\hat{H}_0 + E_L \delta_e(t)\ \hat{\delta},
    \label{eq:H_fluxonium_time}
\end{equation}
where the classical drive field $\delta_e(t)$ couples to the quantum phase operator $\hat{\delta}$. At the sweet-spot, this operator only connects states of different parity. Bang amplitude is parametrized with an effective TLS amplitude, satisfying $\phi(t) \hat{\sigma}_x = 2 E_L \delta_x \delta_e(t) \hat{\sigma}_x$, where $\delta_x=\langle 1| \hat{\delta}|0\rangle$. Executing the time-optimal bang-bang sequences derived in Section II demands large effective drive amplitudes, which applied to a multilevel system inevitably excites transitions outside the computational subspace, threatening to degrade gate fidelity through leakage.

\begin{figure}[t!]
    \centering
\includegraphics[width=0.5\textwidth]{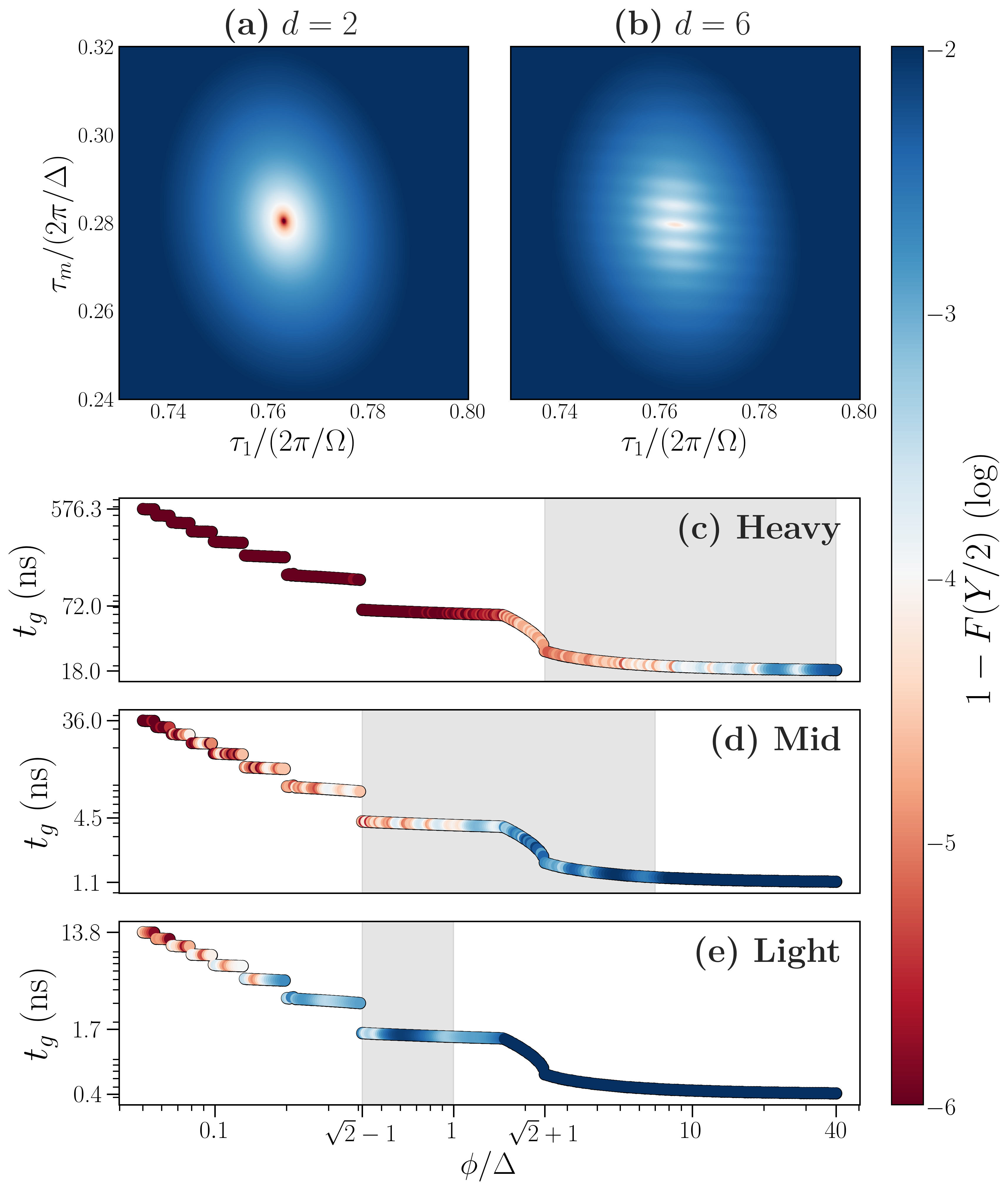}
    \caption{\textbf{Effect of leakage in the optimization.} (a) Color map of the $Y/2$ gate infidelity, $\log_{10}(1-F)$ as a function of $\tau_1/(2\pi/\Omega)$ and $\tau_m/(2\pi/\Delta)$ obtained for the unitary evolution of the Heavy fluxonium truncated to a two-level system and (b) for the full Hamiltonian, at a fixed drive amplitude of $\phi/\Delta = 20$ and $N=2$. (c-e) Optimized $Y/2$ total absolute gate time $t_g$ as a function of the effective transverse drive amplitude $\phi/\Delta$ across the Heavy, Mid and Light regimes of the fluxonium. The color of each point indicates the resulting fidelity. The shaded area indicates the amplitude region of interest for each fluxonium regime for the remainder of this work.}
    \label{fig:fig4}
\end{figure}

\subsection{Coherent suppression of leakage}
Upon the abrupt application of a strong bang, the initial state decomposes into the eigenbasis of the instantaneous Hamiltonian. The resulting phase accumulations manifest as population leakage in the computational basis. For multilevel systems, average gate fidelity is generalized to \cite{wood2018quantification}
\begin{equation}
    F(U, V) = \frac{\frac{1}{d} |\operatorname{tr}(U_q^\dagger V)|^2 + (1 - L_1)}{(d+1)}
    \label{eq:fid_multilevel}
\end{equation}
where the leakage rate $L_1$ for a given protocol is defined as,
\begin{equation}
    L_1(t) = 1 - \frac{1}{2}\operatorname{tr}\left( U_q^\dagger(t) U_q(t)\right),
    \label{eq:leakage_def}
\end{equation} 
and $U_q$ is the time evolution operator truncated to the qubit subspace, extracted from the total unitary evolution of the multilevel system. 
Note that for a TLS, Eq. \eqref{eq:fid_multilevel} trivially reduces to Eq. \eqref{eq:fidelity_TLS}, as the leakage rate is $L_1=0$.

In Fig. \ref{fig:fig3}(e,f), we present maps of the leakage rate $L_1$ in logarithmic scale as a function of the normalized bang durations $\tau_1 / (2\pi/\Omega)$ and $\tau_m / (2\pi/\Omega)$ for the $Y/2$ and $X/2$-gate protocols, respectively. Despite the complexity of the multilevel dynamics, the resulting leakage landscape exhibits a structure governed by transitions between $|0\rangle \leftrightarrow |3\rangle$ and $|1\rangle \leftrightarrow |2\rangle$, with negligible dynamics occurring directly between leakage states, as indicated in Fig. \ref{fig:fig3} by the colored lines. This is a direct consequence of the selection rules for the matrix elements of $\hat{\delta}$. We find that under some drive conditions, these isolated channels exhibit complete population return to the computational subspace, which we call {\it coherent suppression of leakage}. We present below exact analytic expressions for such conditions valid in the TLS limit. There are some special cases where they can be simultaneously satisfied for the dominant channels and most of gate leakage can be effectively eliminated. 

We model $|0\rangle \leftrightarrow |3\rangle$ and $|1\rangle \leftrightarrow |2\rangle$ transitions as independent TLS parameterized by transition frequencies $\Delta_{ij}$ and effective Rabi frequencies $\Omega_{ij}= \sqrt{\Delta_{ij}^2 + \phi^2}$. In App. \ref{ap:gate_analytics}.3 we derive the conditions for leakage suppression for the $Y/2$ and $X/2$ protocols. For the bang-idle-bang protocol (see Fig. \ref{fig:fig3}(e)), two distinct suppression conditions emerge. The first one is 
\begin{equation}
    \tau_1 = k\frac{2\pi}{\Omega_{ij}}, \quad k=1,2,\dots
\end{equation}
which makes leakage for each channel vanish exactly at each bang, producing the vertical lines in Fig. \ref{fig:fig3}(e). 
The second condition is
\begin{equation}
    \tan\left(\frac{\Delta_{ij}\tau_m}{2}\right) = -\frac{\Delta_{ij}}{\Omega_{ij}}\tan\left(\frac{\Omega_{ij}, \tau_1}{2}\right)
\end{equation}
which in the high-anharmonicity limit ($\Delta_{ij} \gg \phi$), simplifies to $\Delta_{ij} (\tau_1+\tau_m) = 2\pi$. This interference condition produces the diagonal lines with negative slope in Fig. \ref{fig:fig3}(e).

For the three-bang protocol (see Fig. \ref{fig:fig3}(f)), coherent suppression of leakage occurs when
\begin{equation}
    \cot\left(\frac{\Omega_{ij} \tau_2}{2}\right) = \frac{\cos(\Omega_{ij} \tau_1) + 4\frac{\Delta_{ij}^2}{\Omega_{ij}^2}\sin^2\left(\frac{\Omega_{ij} \tau_1}{2}\right)}{\sin(\Omega_{ij} \tau_1)}
\end{equation}
which manifests as the oscillatory contours in Fig. \ref{fig:fig3}(f). For both protocols, these approximate expressions work very well and allow to anticipate the existence of privileged pulse sequences which minimize leakage error and maximize fidelity.  

\subsection{Numerical optimization with the full multilevel circuit}
The optimization algorithm is sensitive to the oscillations of the leakage error that occur as a function of $\tau_1$ and $\tau_m$ (see Fig. \ref{fig:fig3}(e, f)).

In Fig. \ref{fig:fig4}(a) we show a fidelity map in log-scale for the $Y/2$ gate protocol in the Heavy regime as a function of $(\tau_1$, $\tau_m)$ obtained for the Hamiltonian Eq. (\ref{eq:H_fluxonium_time}) truncated to two-levels. Maps of this type were used to obtain the TLS results presented in Fig. \ref{fig:fig2}. In Fig. \ref{fig:fig4}(b) the same protocol is applied to the full fluxonium circuit. In practice we calculate the time evolution projecting the Hamiltonian onto the undriven eigenbasis, considering the lowest six levels, which we have verified to be accurate enough. The fidelity map in Fig. \ref{fig:fig4}(b) shows oscillations as a function of $(\tau_1$, $\tau_m)$ which are consistent with the behavior presented in Fig. \ref{fig:fig3}(e) for the leakage error. The optimization sweeps across these infidelity valleys produced by coherent suppression leakage and look for the fastest highest-fidelity protocol.

In Figs. \ref{fig:fig4}(c-e) we present the optimized gate time $t_g$ for the $Y/2$ protocol as a function of the bang amplitude $\phi/\Delta$ across the different fluxonium regimes introduced in Fig. \ref{fig:fig3}. While the absolute gate time $t_g$ decreases for higher qubit frequencies, the corresponding lower anharmonicities limit the fidelity for stronger drives due to leakage. Furthermore, Figs. \ref{fig:fig4}(c-e) reveal fluctuations in fidelity as a function of $\phi/\Delta$. These high-fidelity solutions emerge when the TLS optimizer finds $(\tau_1, \tau_m)$ values (the red minimum in Fig. \ref{fig:fig4}(a)) that align with a leakage suppression valley (Fig. \ref{fig:fig4}(b)).

Ultimately, as we explore in the next section, the general impact of leakage can be mitigated by applying continuous waveform smoothing. This produces a more adiabatic evolution, deepening the leakage valleys and expanding the available parameter space for high-fidelity control, albeit increasing $t_g$.

\section{Physical Gate Synthesis and Benchmarking}

\begin{figure*}[t!]
    \centering
\includegraphics[width=\textwidth]{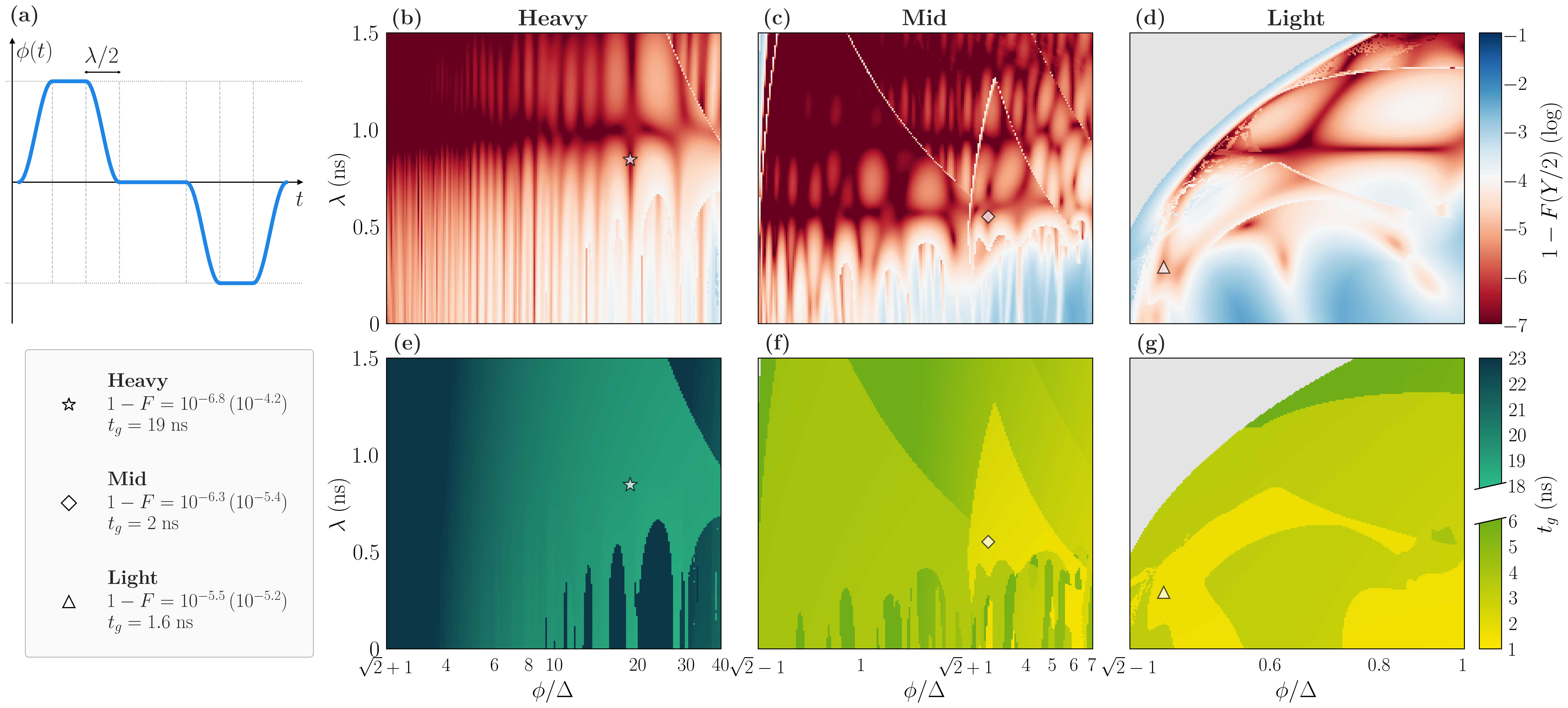}
    \caption{\textbf{Smoothing and fluxonium $Y/2$ gates.} (a) Schematic of the pulse smoothing protocol. Cosine ramps of width $\lambda/2$ are added at the bang transitions, increasing the total gate time by $2\lambda$. (b-d) Optimized $Y/2$ gate infidelity, $\log_{10}(1-F)$, as a function of the effective transverse drive amplitude $\phi/\Delta$ and the smoothing parameter $\lambda$ for the Heavy, Mid, and Light fluxonium regimes. High-fidelity solutions arise where the target coherent rotation coincides with the leakage suppression channels. (e-g) Corresponding optimal gate times $t_g$. The numerical optimizer searches for the fastest protocol achieving an infidelity below $1-F \leq 10^{-4}$; if this threshold is unreachable, it converges to the lowest possible error. \textbf{Table:} Final benchmark metrics obtained via unconstrained optimization of all pulse parameters, seeded by the points marked with the corresponding symbols. Fidelities are reported for closed (open) system simulations.}
    \label{fig:fig5}
\end{figure*}

\subsection{Smoothing}
In Fig. \ref{fig:fig5} (a) we present our strategy to smoothen the sharp bang-bang pulses. We append cosine ramps of temporal width $\lambda/2$ at the transitions between idling segments and bangs. This parametrization allows easy control of the duration added to the total gate time since $N$ bang pulses would add $N\times\lambda$ time to the sequence. However, as we show next, this is not as time expensive as it seems. The reason is that the evolution occurring during the ramp can be also harnessed to reduce the duration $(\tau_1,\tau_m)$ of the bangs themselves, as we show in what follows.

For the general case we employ a two-step optimization strategy: first, we perform a contour sweep across the parameter space $(\tau_1, \tau_m)$ using a truncated two-level Hamiltonian (see Fig. \ref{fig:fig4}(a)), in order to find the non-leakage solution with smoothing. Then we perform a new sweep with these solutions as seeds, using the full multilevel Hamiltonian, to fine-tune and reduce leakage. For a given $\lambda$, smoothing shifts the $(\tau_1,\tau_m)$ coordinates of the perfect fidelity basins.

\subsection{Dissipation}
In the spirit of presenting realistic metrics that could be compared with the fidelities and gate times reported in the literature \cite{zhang2021universal, rower2024suppressing, ferreyra2026optimizing, nguyen2022blueprint}, we discuss the effect of dissipation in particular gates determined after the optimization. Open-system simulations were done using the Universal Lindblad Equation (ULE), which, unlike standard approaches, does not require a secular approximation \cite{nathan2020universal}. Furthermore, to include non-Markovian $1/f$ flux noise, we average the open-system propagators over static flux offsets sampled via Gauss-Hermite quadrature (see the details in App. \ref{ap:open_system}) \cite{ithier2005decoherence, townsend2016fast}.

\subsection{The $\mathbf{Y/2}$ Gate}
We start discussing the full optimization of the bang-idle-bang protocol (see Fig. \ref{fig:fig5}(a)) for the implementation of $Y/2$ gates. We focus on the ranges of $\phi/\Delta$ shown in Fig. \ref{fig:fig4}(c-e). Although slower than the $X/2$ gate, the time-optimal $Y/2$ protocol offers advantages: it effectively acts as a spin-echo sequence, suppressing first-order $1/f$ flux noise \cite{vandersypen2004nmr, hahn1950spin}. Furthermore, because the full pulse possesses a net-zero area, it mitigates long-term distortions in experimental flux lines \cite{rol2019fast}.

In Fig. \ref{fig:fig5}(b-d) we present the fidelities in log-scale and in Fig. \ref{fig:fig5}(e-g) the total gate time for the optimal $Y/2$ protocol across the Heavy, Mid, and Light fluxonium regimes. Each point represents the result of the optimization over $(\tau_1$, $\tau_m)$, and the first-bang sign, with fixed $\phi$ and $\lambda$. The fidelity maps exhibit a nontrivial structure, with high-fidelity solutions emerging precisely where the two-level gate fidelity coincides with multi-channel coherent suppression of leakage. The color-scale is chosen such that white corresponds to 10$^{-4}$ error, which is desirable in experiments, red is better and blue is worse. In the range of $\phi/\Delta$ where the $N=2$ bang sequence is optimal, the fidelity of the strong-drive protocol worsens as the fluxonium regime moves from Heavy to Light, mostly due to the lower anharmonicity. Remarkably, introducing sub-nanosecond smoothing (increasing $\lambda$) drastically improves fidelity while preserving near-optimal gate speeds.

As expected, the optimized maps (Fig. \ref{fig:fig5}(e-g)) reveal a general decrease in $t_g$ with increasing amplitude. The amplitude regions with fast solutions broaden with even a small $\lambda$. For example, for $\lambda=0.8$\ ns there are fast solutions in the three fluxonium regimes. In the Heavy fluxonium regime, total gate times saturate near $18\;\text{ns}$, closely approaching the theoretical bound of $\tau_L/4 \sim 17.5\;\text{ns}$. As we transition to lighter fluxonium architectures, the velocities $\Omega = \Delta \sqrt{1 + (\phi/\Delta)^2}$ increase, scaling down gate times. Across all regimes, the optimal rise and bang durations fall within the $0.2-0.35\;\text{ns}$ range. These timescales are well within the analog bandwidths and sampling rates of high-performance commercial AWGs \cite{keysight_awg, tektronix_awg}, which routinely operate at $50-65\;\text{GSa/s}$. For experimental setups constrained by lower sampling rates, smaller drive amplitudes can be selected while still maintaining competitive gate speeds.

Ultimately, we evaluate the performance of these synthesized protocols in terms of gate time . To ensure a fair comparison across different control schemes, we benchmark our sequences using the gate time $\overline{t_g}$ averaged over the standard 24-gate Clifford set, as conventionally used in randomized benchmarking to determine experimental gate fidelity \cite{knill2008randomized} (see App. \ref{sec:fair_comparisons}). Unlike resonant microwave protocols, which implement instantaneous virtual-$Z$ in software \cite{mckay2017efficient}, lab-frame control requires finite idling intervals to perform $Z$-rotations \cite{zhang2021universal}. The comparison between both protocols in the different fluxonium regimes is presented in Table \ref{tab:Y_2_results}. Here, $\tau_L$ is the Larmor period, $\overline{t_g}$ is the average gate time and $t_{Y/2}$ the time of the $Y/2$ gate, for each control scheme. To establish a rigorous baseline, the resonant control scheme utilizes a standard DRAG-optimized cosine pulse. Because virtual-$Z$ gates lose fidelity due to counter-rotating terms under strong driving, the pulse duration $t_{Y/2}$ is chosen such that closed-system virtual-$Z$ errors remain below $10^{-5}$. The results in Table \ref{tab:Y_2_results} demonstrate that even with the penalization of long idling times, bang-bang protocols consistently outperform resonant microwave control and lab-frame Landau-Zener-Stückelberg (LZS) gates \cite{ferreyra2026optimizing} across all tested regimes. We highlight the comparison against the LZS scheme, driven by a single-period sinusoidal pulse, as it represents a conceptually similar lab-frame approach. In the Light regime, our protocol yields a marginal improvement over LZS in average gate time. However, this advantage scales favorably for heavier fluxoniums, where slower intrinsic system dynamics bottleneck the LZS sweeps. In the Heavy regime, our $Y/2$ protocol outperforms LZS by over 16 ns and recovers the best approach established in the literature, which we call the triangular bangs protocol \cite{zhang2021universal}. Using the device parameters from this reference, our approach accelerates the average gate time by 2 ns. Furthermore, in the Mid regime, we additionally evaluate our protocol against a commensurate pulse scheme \cite{rower2024suppressing}, which exhibits a significantly longer average gate time due to its specific Clifford decomposition (App. \ref{sec:fair_comparisons}).

\begin{table}[htbp]
    \centering
    \renewcommand{\arraystretch}{1.3}
    \resizebox{\columnwidth}{!}{%
    \begin{tabular}{l l l c c c} 
        \toprule
        \textbf{Regime} & $\tau_L$ \textbf{(ns)} & \textbf{$Y/2$ Control Scheme} & $t_{\pi/2}$ \textbf{(ns)} & $\overline{t_g}$ \textbf{(ns)} & \textbf{Ref.} \\
        \midrule
        \multirow{4}{*}{\textbf{Heavy}} & \multirow{4}{*}{72} 
                                        & \textbf{Bang-Bang} & \textbf{19} & \textbf{50.7} & - \\
        &                               & Triangular bangs   & 21          & 52.7          & \cite{zhang2021universal} \\
        &                               & LZS   & 35.3          & 67         & - \\
        &                               & Resonant\cite{footnote_resonant}          & 108          & 108          & - \\
        \midrule
        \multirow{5}{*}{\textbf{Mid}}   & \multirow{5}{*}{4.5} 
                                        & \textbf{Bang-Bang}   & \textbf{2} & \textbf{3.8} & - \\
        &                               & LZS                  & 2.3        & 4.3           & - \\
        &                               & Resonant             & 6.75          & 6.75           & - \\
        &                               & Commensurate ($2\tau_L$) & 9        & 24.8\cite{footnote_commensurate}           & \cite{rower2024suppressing} \\
        &                               & Commensurate ($1\tau_L$) & 4.5          & 12.4           & \cite{rower2024suppressing} \\
        \midrule
        \multirow{3}{*}{\textbf{Light}} & \multirow{3}{*}{1.7} 
                                        & \textbf{Bang-Bang} & \textbf{1.6} & \textbf{2.3} & - \\
        &                               & LZS                & 1.9          & 2.7            & \cite{ferreyra2026optimizing}  \\
        &                               & Resonant           & 3.2           & 3.2           & - \\
        \bottomrule
    \end{tabular}}%

    \caption{Benchmarking of average Clifford gate times ($\overline{t_g}$) across different fluxonium regimes. Our proposed lab-frame bang-bang sequence is compared against both generalized resonant control and regime-specific state-of-the-art literature. Here, $\tau_L$ is the Larmor period and $t_{\pi/2}$ is the physical duration of the required $\pi/2$ rotation. Despite the finite idling times required for $Z$-rotations in bang-bang control, the proposed protocol systematically outperforms both resonant and commensurate baselines.}
    \label{tab:Y_2_results}
\end{table}

\subsection{The $\mathbf{X/2}$ Gate}
While the $Y/2$ protocol reduces gate times across most configurations, in the case of Heavy fluxonium, we approach the theoretical speed limit $t_g \ge \tau_L/4\sim 17.5\; \text{ns}$. Further accelerating Heavy fluxonium control would require implementing an $X/2$ which involves non net-zero pulses: this asymmetric protocol exposes the qubit to un-echoed $1/f$ flux noise and long-term distortions in the flux line. Fortunately, these distortions in the control lines can be fought using pulse pre-distortion techniques, and emerging cold on-chip electronics that promise to bypass room-temperature artifacts entirely \cite{herr2011ultra, Przybysz2026RQL, hellings2025calibrating, rol2020time}. Motivated by these engineering solutions, we explore the temporal advantages of the $X/2$ protocol in the Heavy regime.

Although the sequence lacks an intrinsic echo, driving the qubit at strong amplitudes saturates the transition frequency derivative, $\frac{\partial \Delta}{\partial \phi}$ (Fig. \ref{fig:fig3}(b)). Because the pure-dephasing time scales as $T_\phi \propto \left| \frac{\partial \Delta}{\partial \phi} \right|^{-1}$ \cite{ithier2005decoherence}, this saturation suppresses the qubit's sensitivity to flux fluctuations. Furthermore, operating at these strong amplitudes reduces the overall execution time $t_g$, minimizing the temporal window for decoherence, maintaining high target fidelities despite the un-echoed noise.

The time-optimal $X/2$ protocols for TLS demand asymptotically short outer bangs (App. \ref{ap:gate_analytics}). We relax the maximum-amplitude constraint ($\phi(t)=\pm\phi_\text{max}$) by treating the amplitude ratio $\mu = \phi_{\text{outer}}/\phi_{\text{middle}}$ as a free optimization parameter. Because the sequence only requires brief initial and final pulses, they can be synthesized at lower amplitudes, demanding less waveform smoothing. The structure of this asymmetric bang-bang protocol parallels sequences proposed for fast state transfer \cite{hsiao2025efficient}. Applying this modified three-bang protocol with $\mu\approx 0.01$ to the Heavy fluxonium regime yields an $X/2$ gate with an open-system infidelity of $1-F = 10^{-5}$ and an execution time of $t_g = 1.9\;\text{ns}$, one order of magnitude faster than the $19\;\text{ns}$ $Y/2$ protocol (Table \ref{tab:Y_2_results}). Even after including the contribution of $Z$-idling times, the $X/2$ protocol achieves an average Clifford time of $\overline{t_g} = 33\;\text{ns}$. Although this average is dominated by the idling periods, it still outperforms the $50\;\text{ns}$ average of the $Y/2$ configuration. 

\subsection{Arbitrary phase $\pi/2$ gates}
 
While our analysis was primarily centered on $X/2$ and $Y/2$ gates, this framework can be naturally extended to arbitrary $\pi/2$ rotations characterized by a continuous phase $\theta$ (where $\theta=0$ and $\theta=\pi/2$ recover the $X/2$ and $Y/2$ gates, respectively). As detailed in App. \ref{ap:gate_analytics}4, these generalized gates in the strong-drive regime are synthesized using either three-bang or bang-idle-bang sequences, but given the lack of global symmetry conditions, the sequences require three independent durations: $\tau_1$, $\tau_2$, and $\tau_3$. The target phase $\theta$ is determined by the temporal asymmetry $|\tau_1 - \tau_3|$. For small $\theta$, the time-optimal solution remains a three-bang protocol. As $\theta$ increases to a critical angle $\theta_c$, $\tau_3\rightarrow 0$, reducing the sequence to two bangs. This critical transition angle is given analytically by:
\begin{equation}
\sin{\theta_c} =  \frac{\sqrt{2}+1}{\phi/\Delta}.
\end{equation}
Increasing $\theta$ beyond this threshold introduces an idling segment between the pulses, smoothly transitioning from $N=3$ to $N=2$.

\section{Conclusions}

In this work, we addressed the fundamental question of identifying the time-optimal waveform for implementing transverse $\pi/2$ quantum gates. First, by applying Optimal Control Theory and Pontryagin's Principle, we rigorously constrained the arbitrary form of the drive pulse to a highly specific family of sequences: bang-bang pulses. Beginning with the general case of a transversally driven two-level system (TLS), we derived the time-optimal bang-bang sequences required to implement $X/2$ and $Y/2$ gates with ideal fidelity. For the TLS, we provided analytic results determining the existence of solutions as a function of the normalized maximum drive amplitude, $\phi/\Delta$, as well as establishing strict lower bounds for the theoretical gate times. 

To test these bounds in a realistic, multilevel low-frequency architecture, we applied the bang-bang protocol to the superconducting fluxonium circuit, which allows us to explore diverse regimes of qubit frequency and anharmonicity. Subjecting a multilevel system to these extreme bang-bang protocols naturally introduces leakage errors. We propose two distinct strategies to mitigate this: coherent suppression of leakage and continuous pulse smoothing. During the optimization of $Y/2$ gates, we observed an oscillatory pattern in the fidelity landscape as a function of the bang durations. A closer analysis revealed that these oscillations correspond to specific sequence timings where phases accumulated during evolution outside the computational subspace destructively interfere to return the population to the qubit subspace. By exploiting these ``error valleys,'' we found fast, high-fidelity gate protocols across the Heavy, Mid, and Light fluxonium regimes. Consequently, due to this coherent interference phenomenon, there is no ``single best operating point'' for a given fluxonium architecture. Alternatively, we demonstrated that introducing a continuous half-cosine ramp to smooth the sharp bang pulses, mimicking realistic arbitrary waveform generators (AWGs) or experimental high-pass filtering, broadens these high-fidelity regions in parameter space. To better compare these fast protocols with experimental implementations, we calculated the open-system fidelity using the Universal Lindblad Equation (ULE) formalism. Furthermore, by incorporating a Gaussian ensemble average to model $1/f$ flux noise, we confirmed that our protocols yield high-fidelity operations that are faster than current experimental implementations and well within reach of modern setups. 

Finally, we evaluated the practical utility of these protocols against the standard resonant paradigm, where virtual-$Z$ gates can be implemented instantaneously in software; an undeniable advantage over lab-frame control, which requires finite idling times. To address this, we benchmarked the protocols using the average gate time computed over the full set of 24 single-qubit Clifford gates. In the Heavy fluxonium regime, while the bang-bang protocol yields a $>5\times$ speedup for specific $\pi/2$ rotations, this advantage is diluted in the average Clifford time due to the slow Larmor rotations required for $Z$ gates. Nonetheless, the bang-bang scheme still provides a roughly $2\times$ improvement over the resonant case overall. In the Mid regime, the bang-bang approach continues to clearly outperform resonant microwave control, reducing the average Clifford time by nearly a factor of two. Surprisingly, even in the Light regime, bang-bang protocols remain highly competitive ($2.3$ ns vs the resonant baseline of $3.2$ ns), as virtual-$Z$ gates cannot be implemented with infinite precision as the required pulse amplitudes become stronger.

To complete our toolkit, we presented two additional extensions. First, we introduced a modified $X/2$ sequence to achieve faster rotations in the Heavy fluxonium regime. Because odd-bang sequences (positive-negative-positive bang) lack idling segments, their execution time is not theoretically bounded. By relaxing the pulse shape to allow an asymmetric central amplitude ($\mu\vert{}\phi_{\text{middle}}\vert{} = \vert{}\phi_{\text{outer}}\vert{}$), we demonstrated that gates with an open-system infidelity of $10^{-5}$ and an execution time of $1.9 \text{ ns}$ can be realized, an order of magnitude faster than the best $Y/2$ protocol. For this configuration, the average Clifford time is reduced to $\overline{t_g} = 33 \text{ ns}$, outperforming the $50 \text{ ns}$ average of the $Y/2$ configuration, despite being dominated by idling periods. Ultimately, alongside our final extension generalizing $\pi/2$ rotations to an arbitrary axis, these results provide a comprehensive framework for synthesizing time-optimal, high-fidelity gates.

\section*{Acknowledgements}
We acknowledge support from CNEA, ANPCyT (PICT 2019-0654), CONICET: PIP 11220200101825CO and PIP
11220220100212CO. 
\appendix
\counterwithin{figure}{section}
\section{Time optimality of protocols}\label{ap:pontryagin}

We consider a transversally driven TLS described by the Hamiltonian $\hat{H}(t) = \frac{\Delta}{2} \hat{\sigma}_z + \frac{\phi(t)}{2}\hat{\sigma}_x$, governed by the Schrödinger equation $\frac{d}{dt}\ket{\psi(t)} = -i \hat{H}(t) \ket{\psi(t)}$. The control field is bounded such that $\phi(t)\in [-\phi_\text{max},+\phi_\text{max}]$. Subject to the boundary condition $\ket{\psi(t_g)} = \ket{\psi_\text{target}}$, we seek to determine the optimal control waveform $\phi(t)$ that minimizes the total gate time ${\cal J} = \int_0^{t_g} 1 \, dt$.

According to Pontryagin's Principle (PP) \cite{boscain2021introduction}, the optimal control field $\phi(t)$ must globally minimize the Pontryagin Hamiltonian, defined as:
\begin{equation}
    \mathcal{H}_{oc}(t) = 1 + \Re[-i\langle p| H(t) |\psi\rangle]
\end{equation}
This result is obtained variationally by minimizing $\cal{J}$, with Schrödinger's equation as a constraint and the dual vector $\bra{p}$ acting as a time-dependent costate (Lagrange multiplier). Because the control field $\phi(t)$ enters $\hat{H}(t)$ linearly, the minimization of $\mathcal{H}_{oc}(t)$ is entirely determined by the sign of the switching function $S_x(t) = \Re(i\langle p|\hat{\sigma}_x|\psi\rangle)$. The PP control conditions are strictly bang-bang:
\begin{itemize}
    \item For $S_x(t) > 0$, the control field saturates to its minimum bound, $\phi(t)=-\phi_\text{max}$.
    \item For $S_x(t) < 0$, the control field saturates to its maximum bound, $\phi(t)=\phi_\text{max}$.
\end{itemize}

If the switching function vanishes over a finite interval ($S_x(t) = 0$), the control field follows a ``singular arc''. To determine the system behavior during these intervals, we generalize the switching function by defining the costate vector $S_j = \Re (i\langle p|\hat{\sigma}_j|\psi\rangle)$. Differentiating respect to time yields the costate equations of motion:
\begin{equation}
    \dot{S}_x = -\Delta S_y, \quad \dot{S}_y = \Delta S_x - \phi S_z, \quad \dot{S}_z = \phi S_y
\end{equation}
During a singular arc, $S_x(t) = 0$ implies $\dot{S}_x = 0$, which strictly requires $S_y(t) = 0$. Consequently, the derivative $\dot{S}_y$ must also vanish, reducing the second equation to $\phi S_z = 0$. Because PP dictates that the costate multiplier $\bra{p}$ can never be identically zero, $S_x$, $S_y$, and $S_z$ cannot simultaneously vanish. Therefore, $S_z \neq 0$, which strictly requires the control field to vanish ($\phi = 0$). Thus, a singular arc exclusively manifests as a zero-amplitude ``idle'' segment.

The parameter space is now restricted to $\phi\in\{\pm \phi_\text{max}, 0\}$. The dynamics of this costate vector determine the shape of the time-optimal protocols. Combining the expressions for $\dot{S}_x$ and $\dot{S}_z$ yields $\dot{S}_x = -\frac{\Delta}{\phi} \dot{S}_z$, leading to the uncoupled equation of motion for $S_x$:
\begin{equation}
    \ddot{S}_x + \Omega^2 S_x = \text{const} \label{eq:Sx_eom}
\end{equation}
Consider the transition from a singular arc (idle) to a bang at an arbitrary intermediate time $t=t_0$. Because the system transitions from a zero-amplitude interval, the boundary conditions are $S_x(t_0) = 0$ and $\dot{S}_x(t_0) = 0$. Solving Eq. (\ref{eq:Sx_eom}) with these conditions straightforwardly yields the general solution $S_x(t) \propto (1 - \cos(\Omega (t-t_0)))$. To transition out of this bang into a new idle segment, $S_x(t)$ must return to zero, which exclusively occurs at intervals $t - t_0 = k \frac{2\pi}{\Omega}$. However, the unitary propagator of such bangs becomes identity, preventing such solutions from minimizing the total gate time.

This demonstrates that an \textit{idle-bang-idle} sequence is strictly suboptimal. Singular arcs may only exist exactly once between two isolated bangs ($N=2$, i.e., \textit{bang-idle-bang}), or adjacent to a single bang ($N=1$, i.e., \textit{idle-bang} or \textit{bang-idle}). Furthermore, as will be demonstrated below, global time symmetries dictated by our target gates discard these asymmetric $N=1$ cases. Consequently, any time-optimal protocol containing more than two bangs ($N \ge 3$) cannot feature idle segments and must be strictly bang-bang. To the best of our knowledge, this necessary constraint on time-optimal pulse synthesis constitutes a novel result.

Furthermore, for protocols where $N \ge 3$, transitions between bangs of opposite signs occur when $S_x=0$ but $\dot{S}_x \neq 0$. Because the switching function behaves as a continuous oscillator with a fixed effective frequency $\Omega$, the condition $S_x(\tau_m) = 0$ enforces a uniform, constant duration $\tau_m$ for all intermediate bangs. Only the durations of the outer bangs ($\tau_1$ and $\tau_N$) are exempt from this strict constraint.

The parameter space can be further reduced by enforcing global time symmetries set by the target gates. Because the Hamiltonian is composed exclusively of symmetric operators ($\hat{\sigma}_z$ and $\hat{\sigma}_x$), any single constant-bang unitary evolution $U(\phi)$ is equal to its transpose, $U^T = U$. 

For an $X/2$ rotation, the target operator is symmetric ($(X/2)^T = X/2$). The synthesized sequence must therefore equal its own reversal:
\begin{equation}
    U_1(\phi_1) \dots U_N(\phi_N) = U_N(\phi_N) \dots U_1(\phi_1)
\end{equation}
A term-by-term comparison yields $U_1(\phi_1) = U_N(\phi_N)$, enforcing the symmetry conditions $\phi_1 = \phi_N$ and $\tau_1 = \tau_N$. Because time-optimal bangs must strictly alternate in sign, the initial and final bangs can only share the same sign if the sequence contains an odd number of bangs ($N=3, 5, \dots$). This structural symmetry requirement naturally discards the \textit{idle-bang} and \textit{bang-idle} solutions.

The target $Y/2$ gate is antisymmetric, meaning $(Y/2)^T = \exp(-i \frac{\pi}{4}\sigma_y)^T=\exp(i \frac{\pi}{4}\sigma_y)$. Applying Pauli conjugation ($\hat{\sigma}_z U(\phi) \hat{\sigma}_z = U(-\phi)$) and using the relation $(Y/2)^T = \hat{\sigma}_z (Y/2) \hat{\sigma}_z$, we equal the time-reversed bangs to their negative-amplitude counterparts:
\begin{equation}
    U_1(\phi_1) \dots U_N(\phi_N) = U_N(-\phi_N) \dots U_1(-\phi_1)
\end{equation}
This indicates that $U_1(\phi_1) = U_N(-\phi_N)$, which dictates that the outer bangs must possess identical durations ($\tau_1 = \tau_N$) but opposite amplitudes ($\phi_1 = -\phi_N$). To respect the alternating sign condition while ending on an inverted sign, the $Y/2$ sequence must necessarily consist of an even number of bangs, ($N=2, 4, \dots$).

Consequently, the optimization space collapses to a highly constrained parameter set: the saturated amplitudes $\pm\phi_{\text{max}}$, the total number of bangs $N$ (restricted to odd integers for $X/2$ gates and even for $Y/2$ gates), the outer bang duration $\tau_1$, and the uniform inner bang duration $\tau_m$.

\section{Numerical optimization for time-optimality}\label{ap:optimization}

To determine the absolute minimum gate durations, we sweep the total gate time $t_g$ in strictly increasing increments, rather than performing an unconstrained optimization over fidelity.

For a fixed drive amplitude $\phi/\Delta$ and bang number $N$, the parameter space is spanned by the outer and inner bang durations, $\tau_1$ and $\tau_m$. The optimization algorithm sweeps this landscape along fixed-time contours (e.g., $t_g = 2\tau_1 + (N-2)\tau_m$ for $N \ge 3$). When a segment of the contour yields an infidelity below a predefined threshold (specifically, $1-F \le 10^{-2}$), the coordinates are recorded as a distinct fidelity basin. As the total time contour expands outward, these basins dynamically merge and drift across the parameter space. To systematically track these drifting basins, we frame their correlation across contours as a combinatorial assignment problem. A Hungarian assignment algorithm correlates high-fidelity intervals across successive contour lines. By penalizing abrupt discontinuities in basin width and center position, this approach prevents the optimizer from losing track of converging solutions. 

Each time a new, isolated basin is identified, a low-iteration gradient descent probes its local minimum. If this local minimum satisfies a stricter infidelity threshold ($1-F \le 10^{-4}$), an L-BFGS algorithm is deployed to extract the exact optimal coordinates. Because the search space is expanded strictly in increasing temporal increments, the algorithm halts immediately upon finding a converged solution, thereby ensuring global time-optimality.

To apply the optimization algorithm to the multilevel fluxonium, we employ a two-step strategy.

The initial global sweep is performed by truncating the Hamiltonian to the lowest two energy levels. In the absence of leakage dynamics, the fidelity landscape is computationally smooth, allowing the contour-tracking algorithm to rapidly isolate the coordinates of the fidelity basins. 

Once a valid two-level minimum is identified, the algorithm defines a bounded region in parameter space around these coordinates. A multi-start L-BFGS algorithm is subsequently deployed with the full multilevel Hamiltonian inside this bounding box. This constraint ensures that the optimizer converges to a solution with minimal leakage while remaining localized near the time-optimal target established in the first stage.

To accelerate the numerical evaluation of the unitary evolution, we implement a matrix caching strategy. During the bang segments, the constant drive amplitude $\phi$ permits exact diagonalization, $H(\phi) = V D V^\dagger$, which consequently enables the highly efficient evaluation of the unitary $U(\tau) = V e^{-i D \tau} V^\dagger$ for an arbitrary pulse duration $\tau$.

When incorporating smoothing ($\lambda$), the evolution becomes explicitly time-dependent and lacks a closed-form analytical solution. Rather than numerically integrating the state vectors for each iterative pulse sequence, we compute and cache the full multilevel unitary (or the $d^2 \times d^2$ open-system propagator in the Liouville space). The optimizer subsequently constructs the total evolution operator by concatenating these precomputed smoothed segments with the diagonalized bang propagators. This composite propagator approach allows the algorithm to rapidly sweep $\tau_1$ and $\tau_m$ without repeatedly integrating the time-dependent equations of motion.

\section{Analytic bang-bang expressions}\label{ap:gate_analytics}

The single-qubit dynamics are governed by the driven Hamiltonian $\hat{H}(t) = \frac{\Delta}{2}\hat{\sigma}_z + \frac{\phi(t)}{2}\hat{\sigma}_x$. During the application of a maximum-amplitude pulse $\phi(t) = \pm \phi$, the state vector precesses around the instantaneous axis with an angular speed $\Omega = \sqrt{\Delta^2 + \phi^2}$. Parametrizing the synthesized evolution operator as $\hat{U}_{seq} = a \hat{I} - i(b_x\hat{\sigma}_x + b_y\hat{\sigma}_y + b_z\hat{\sigma}_z)$, the realization of an ideal target gate requires strict constraints on the coefficients.

\subsection{The $Y/2$ Gate}

The bang-idle-bang sequence comprises two saturated pulses of equal duration $\tau_1$ and opposite sign, separated by an idling segment of duration $\tau_m$. The resulting propagator is $\hat{U}_{BIB} = \hat{B}_{-s}(\tau_1) \hat{Z}(\tau_m) \hat{B}_{s}(\tau_1)$, where $s \in \{+1, -1\}$ determines the initial pulse sign. Evaluating this sequence yields the matrix components:

\begin{align}
    a &= \left(\frac{|\phi|^2}{\Omega^2} + \frac{\Delta^2}{\Omega^2}\cos 2x\right)\cos\frac{\gamma}{2} - \frac{\Delta}{\Omega}\sin 2x \sin\frac{\gamma}{2}, \label{eq:a_first}\\
    b_y &= s\left[ -\frac{|\phi|}{\Omega}\sin 2x \sin\frac{\gamma}{2} - 2\frac{|\phi|\Delta}{\Omega^2}\sin^2 x \cos\frac{\gamma}{2} \right], \label{eq:by_Y2}\\
    b_z &= \cos 2x \sin\frac{\gamma}{2} + \frac{\Delta}{\Omega}\sin 2x \cos\frac{\gamma}{2},
\end{align}
where we have defined the dimensionless phases $x=\Omega \tau_1/2$ and $\gamma=\Delta \tau_m$. Synthesizing an ideal target gate $\hat{Y}/2 = \frac{1}{\sqrt{2}}(\hat{I} - i\hat{\sigma}_y)$ requires the enforcement of the conditions $b_x = b_z = 0$ and $a = b_y = \pm 1/\sqrt{2}$. The even symmetry of the pulse sequence intrinsically satisfies $b_x = 0$. Imposing the constraint $b_z = 0$ isolates the idling duration:
\begin{equation}
    \tan\left(\frac{\Delta\tau_m}{2}\right) = -\frac{\Delta}{\Omega}\tan(\Omega \tau_1)
    \label{eq:bsb_bz}
\end{equation}

By resolving the corresponding trigonometric components and substituting them into the condition $a^2 = 1/2$ in Eq. \eqref{eq:a_first}, the system reduces to a polynomial in terms of the variable $y = \sin^2(\Omega\tau_1/2)$. Introducing the dimensionless drive ratio $r = \phi/\Delta$, the trace constraint takes the explicit form:
\begin{equation}
    4r^2(r^2-1)y^2 - 4r^2(r^2+1)y + (r^2+1)^2 = 0.
\end{equation}

This quadratic equation yields two physical branches, whose real-valued solutions ($y \le 1$) are restricted to specific drive amplitude regimes:
\begin{itemize}
\item Branch 1 ($s=+1$): Exists for $\phi/\Delta \ge \sqrt{2} - 1$, yielding the solution $y_1 = \frac{r^2+1}{2r(r+1)}$.
\item Branch 2 ($s=-1$): Exists for $\phi/\Delta \ge \sqrt{2} + 1$, yielding the solution $y_2 = \frac{r^2+1}{2r(r-1)}$.
\end{itemize}

In the regime where Branch 2 is real, it provides the globally optimal gate time $t_g = 2\tau_1 + \tau_m$. Thus, the required pulse duration for the $Y/2$ gate is given by:
\begin{equation}
    \tau_1 = \frac{2}{\Omega} \arcsin\left( \sqrt{ \frac{(\phi/\Delta)^2+1}{2(\phi/\Delta)(\phi/\Delta \pm 1)} } \right)
\end{equation}
where the positive and negative signs correspond to Branch 1 and Branch 2, respectively. The required idling time $\tau_m$ is determined by extracting the smallest positive root of Eq. ~(\ref{eq:bsb_bz}).

Taking the strong-drive limit ($r \gg 1$) for Branch 2, the pulse parameter converges to $y_2 \approx \frac{1}{2}(1 + \Delta/\phi)$. This implies a phase accumulation of $\Omega\tau_1 \approx \pi/2 + \Delta/\phi$, yielding an asymptotic pulse duration of $\tau_1 \approx \pi/(2\phi)$. Substituting this limit into the idling constraint (Eq. ~\ref{eq:bsb_bz}) yields
\begin{equation}
    \tan\left(\frac{\Delta\tau_m}{2}\right) \approx -\frac{\Delta}{\phi} \tan\left(\frac{\pi}{2} + \frac{\Delta}{\phi}\right) = 1
\end{equation}
This fixes the optimal idling duration at exactly one quarter of the Larmor period, $\tau_m = \pi/(2\Delta) = \tau_L/4$. The total asymptotic execution time for the $Y/2$ protocol evaluates to
\begin{equation}
    t_g \approx \frac{\tau_L}{4} + \frac{\pi}{\phi},\quad \left(\frac{\phi}{\Delta}\rightarrow\infty\right)
\end{equation}

\subsection{The $X/2$ Gate}

For a transverse rotation around the $X$-axis, the target unitary is $\hat{X}/2 = \frac{1}{\sqrt{2}}(\hat{I} - i\hat{\sigma}_x)$. Because a perfect-fidelity $\hat{X}/2$ gate cannot be synthesized via a single pulse due to static $Z$-drift, we employ a symmetric three-bang sequence: $\hat{U}_{3B} = \hat{B}_s(\tau_1) \hat{B}_{-s}(\tau_2) \hat{B}_s(\tau_1)$, where $s \in \{+1, -1\}$. Expanding the sequence yields the matrix components:
\begin{align}
    a &= \cos w \cos 2x - \sin w \sin 2x \frac{\Delta^2 - \phi^2}{\Omega^2} \\
    b_z &= \frac{\Delta}{\Omega}\left[ \sin(2x+w) + 4\frac{\phi^2}{\Omega^2}\sin^2 x \sin w \right] \\
    b_x &= \frac{\eta\phi}{\Omega}\left[ \sin(2x-w) - 4\frac{\Delta^2}{\Omega^2}\sin^2 x \sin w \right] \label{eq:bx_X2}
\end{align}

where $x=\Omega \tau_1/2$ and $w=\Omega \tau_2/2$. Achieving unit fidelity against the target state imposes the conditions $b_y=b_z=0$ and $a=b_x=1/\sqrt{2}$. As in the preceding case, the global odd symmetry of the waveform ensures $b_y = 0$. Imposing the constraint $b_z = 0$ dictates that the central pulse must compensate for the $Z$-phase accumulated during the outer pulses:
\begin{equation}
    \cot\left(\frac{\Omega\tau_2}{2}\right) = -\frac{\cos(\Omega\tau_1) + 4\beta^2\sin^2\left(\frac{\Omega\tau_1}{2}\right)}{\sin(\Omega\tau_1)},
    \label{eq:bbb_cot}
\end{equation}

where we defined $\beta=|\phi|/\Omega$. Substituting the trigonometric components of $\Omega \tau_2$ into the trace constraint $a^2 = 1/2$ and simplifying the resulting expression yields the algebraic condition for the outer pulse durations, parameterized by $y = \sin^2(\Omega\tau_1/2)$:
\begin{equation}
    16\beta^2\left(1 + \beta^2\right)y^2 - 24\beta^2y + 1 = 0.
\end{equation}

The existence of real-valued roots strictly dictates a lower bound on the drive amplitude:
\begin{equation}
    \frac{\phi}{\Delta} \ge \frac{1}{\sqrt{7}}.
\end{equation}

Defining the drive ratio $\beta = \phi/\Omega$, the time-optimal outer bang durations are given by
\begin{equation}
    \tau_1 = \frac{2}{\Omega} \arcsin\left( \sqrt{ \frac{3\beta^2 \pm \beta\sqrt{8\beta^2 - 1}}{4\beta^2(1+\beta^2)} } \right),
\end{equation}

where the negative branch yields the time-optimal solution. The total gate time is then $t_g = 2\tau_1 + \tau_2$.

In the asymptotic strong-drive regime ($\phi \gg \Delta$, such that $\beta \to 1$), the negative branch converges to a constant phase, $y_- \to \frac{3-\sqrt{7}}{8}$. Correspondingly, evaluating Eq. ~(\ref{eq:bbb_cot}) demonstrates that the central pulse converges to $\Omega\tau_2/2 \to \pi - \arctan(1/\sqrt{7})$. 

Applying the approximation $\Omega \approx \phi$, the total accumulated phase reduces to:
\begin{align}
    \frac{\Omega t_g}{2} &= 2\arcsin\left(\sqrt{\frac{3-\sqrt{7}}{8}}\right) + \pi - \arctan\left(\frac{1}{\sqrt{7}}\right) \nonumber\\
    & = \pi + \arctan(8 - 3\sqrt{7}).
\end{align}
Ultimately, the absolute theoretical minimum time for the $X/2$ protocol exhibits an inverse scaling with the drive amplitude:
\begin{equation}
    t_g \approx \frac{2}{\phi} \left[ \pi + \arctan(8 - 3\sqrt{7}) \right] \approx \frac{6.408}{\phi}.
\end{equation}

\subsection{Coherent Suppression of Leakage}
We model each leakage channel as an independent TLS, denoted $L$, spanned by a computational state $|i\rangle_L$ and a leakage state $|j\rangle_L$. This subsystem is characterized by its transition frequency $\Delta_L = \Delta_{ij}$ and an effective Rabi frequency $\Omega_L = \sqrt{\Delta_L^2 + \phi^2}$.

To achieve coherent suppression of leakage, the net population transfer to the non-computational subspace must vanish:
\begin{equation}
    P_\text{leak} = |\bra{1_L}U_L\ket{0_L}|^2 = b^2_{x,L} + b^2_{y,L} = 0
\end{equation}

During a $Y/2$ gate, the symmetry of the bang-idle-bang protocol analytically guarantees that $b_{x,L}=0$ for the leakage subspace. Consequently, total leakage suppression is dictated entirely by setting $b_{y,L}=0$, in Eq. \eqref{eq:by_Y2}. Using the effective parameters ($\Delta_L, \Omega_L$), this equation presents two roots corresponding to two distinct conditions for leakage suppression.

The first condition corresponds to single-pulse suppression, where population transfer is fully mitigated within the duration of the individual pulses, rendering the total leakage independent of the idling time:
\begin{equation}
    \tau_1 = k\frac{2\pi}{\Omega_L}, \quad k \in \mathbb{N}.
\end{equation}

The second condition establishes a relationship between the pulse and idling durations:
\begin{equation}
    \tan\left(\frac{\Delta_L \tau_m}{2}\right) = -\frac{\Delta_L}{\Omega_L}\tan\left(\frac{\Omega_L \tau_1}{2}\right).
    \label{eq:y2_leakage_interference}
\end{equation}

In the limit of highly anharmonic architectures, the Rabi frequency simplifies to $\Omega_L \approx \Delta_L$. Inserting this limit into Eq. ~\eqref{eq:y2_leakage_interference} simplifies the condition to:
\begin{equation}
    \Omega_L(\tau_1 + \tau_m) = 2\pi
\end{equation}

Conversely, to suppress leakage during the $X/2$ gate, we evaluate the three-bang sequence acting on the leakage TLS. For this protocol, the waveform symmetry ensures $b_{y,L}=0$. Consequently, coherent suppression of leakage relies entirely on enforcing $b_{x,L}=0$ in Eq. \eqref{eq:bx_X2}, which establishes the corresponding condition:

\begin{equation}
    \cot\left(\frac{\Omega_L \tau_2}{2}\right) = \frac{\cos(\Omega_L \tau_1) + 4\frac{\Delta_L^2}{\Omega_L^2}\sin^2\left(\frac{\Omega_L \tau_1}{2}\right)}{\sin(\Omega_L \tau_1)}
\end{equation}

In the high-anharmonicity limit ($\phi \ll \Delta_{L}$, such that $\Omega_L \approx \Delta_L$), the numerator reduces analytically to $1 + 2\sin^2(\Delta_L \tau_1 / 2) = 2 - \cos(\Delta_L \tau_1)$. This yields the asymptotic suppression condition:
\begin{equation}
    \cot\left(\frac{\Delta_L \tau_2}{2}\right) \approx \frac{2 - \cos(\Delta_L \tau_1)}{\sin(\Delta_L \tau_1)}
\end{equation}

\subsection{Arbitrary phase $\theta$ gate}

The execution speed of lab-frame protocols for low-frequency architectures is fundamentally bottlenecked by the extended durations required for $Z(\theta)$ rotations. This temporal limitation could be mitigated by calibrating generalized transverse $\pi/2$ gates that natively incorporate an arbitrary azimuthal phase $\theta\in[0,\pi/2]$. This phase determines the rotation axis in the equatorial plane, where $\theta = 0$ corresponds to an $X/2$ gate, $\theta = \pi/2$ to a $Y/2$ gate, and values in between yield intermediate $\pi/2$ rotations. Consequently, for small phases $\theta$, the target rotation closely approximates the $X/2$ gate, and the optimal synthesis retains its characteristic three-pulse structure but requires the relaxation of its global temporal symmetry.

The unitary constraints for an arbitrary-phase rotation are defined by:
\begin{equation}
    a = \frac{1}{\sqrt{2}}, \quad b_z = 0, \quad b_x = \frac{\cos\theta}{\sqrt{2}}, \quad b_y = \frac{\sin\theta}{\sqrt{2}}
\end{equation}

By permitting asymmetric outer pulse durations ($\tau_1 \neq \tau_3$) and defining the phase accumulations $x_i = \frac{\Omega\tau_i}{2}$, the three-pulse propagator expands to:
\begin{align}
    a &= \cos(x_1+x_2+x_3) + 2\frac{\phi^2}{\Omega^2} \sin x_2 \sin(x_1+x_3) \\
    b_z &= \frac{\Delta}{\Omega} \left[ \sin(x_1+x_2+x_3) + 4\frac{\phi^2}{\Omega^2} \sin x_1 \sin x_2 \sin x_3 \right] \\
    b_y &= 2\eta \frac{\phi\Delta}{\Omega^2} \sin x_2 \sin(x_1 - x_3) \\
    b_x &= \frac{\eta\phi}{\Omega} \left[ \sin(x_1+x_3)\cos x_2 - \cos(x_1-x_3)\sin x_2 \right.\nonumber \\
    &\quad \left. + 2\frac{\phi^2-\Delta^2}{\Omega^2} \sin x_1 \sin x_2 \sin x_3 \right]
\end{align}

We introduce sum and difference variables for the outer bangs: $\Sigma = x_1 + x_3$ and $\delta = x_1 - x_3$. Imposing the constraint $b_z = 0$ provides an explicit relation for the asymmetry parameter:
\begin{equation}
    \cos\delta = \cos\Sigma - \frac{\Omega^2}{2\phi^2} \frac{\sin(\Sigma+x_2)}{\sin x_2}
\end{equation}

Substituting $\Sigma$ into the trace constraint ($a = 1/\sqrt{2}$) decouples the rotation angle from the asymmetry $\delta$:
\begin{equation}
    \cos(\Sigma + x_2) + 2\frac{\phi^2}{\Omega^2} \sin x_2 \sin\Sigma = \frac{1}{\sqrt{2}}
\end{equation}

Consequently, the trace, and thus the effective rotation angle, is governed by the aggregate outer duration $\Sigma$ and the central pulse $x_2$. Conversely, $b_z$ is compensated by the difference $\delta$, which simultaneously acts as the generator for the target phase $\theta$:
\begin{equation}
    b_y = 2\eta \frac{\phi\Delta}{\Omega^2} \sin x_2 \sin\delta = \eta\frac{\sin\theta}{\sqrt{2}}
\end{equation}

As $\theta$ increases, the required asymmetry $\delta$ scales proportionally, causing the final pulse duration $x_3$ to vanish. At a critical target phase $\theta_c$, this terminal duration evaluates to zero ($x_3 = 0$). Evaluating the condition $x_3 = 0$ under the constraint $b_z=0$ mandates that $x_1 = \pi - x_2$. Substituting this relation into the trace constraint yields:
\begin{equation}
    \sin^2 x_2 = \frac{\Omega^2}{2\phi^2} \left(1 + \frac{1}{\sqrt{2}}\right)
\end{equation}

Evaluating the corresponding expression for $b_x$ determines the critical phase:
\begin{equation}
    \theta_c = \arcsin\left( \frac{\Delta}{\phi} (\sqrt{2} + 1) \right)
    \label{eq:theta_c}
\end{equation}

Operating beyond this critical phase ($\theta > \theta_c$) requires a transition to the bang-idle-bang protocol. Enforcing the constraints $b_z = 0$ and $a = 1/\sqrt{2}$ upon the composite unitary $\hat{U}_{BIB} = \hat{U}_3 \hat{Z} \hat{U}_1$ yields:
\begin{align}
    0 &= \cos\left(\frac{\phi}{2}\right) \sin(x_1 + x_3) - \frac{\Delta}{\Omega} \sin\left(\frac{\phi}{2}\right) \sin(x_1 - x_3) \label{eq:bsb_bz_arb} \\
    \frac{1}{\sqrt{2}} &= \cos\left(\frac{\phi}{2}\right)\cos(x_1 + x_3) - \frac{\Delta}{\Omega} \sin\left(\frac{\phi}{2}\right)\sin(x_1 - x_3) \nonumber\\
    &\quad\quad\quad+ 2\frac{\phi^2}{\Omega^2} \cos\left(\frac{\phi}{2}\right) \sin x_1 \sin x_3 \label{eq:bsb_a_arb}
\end{align}

Finally, the arbitrary phase is geometrically determined by the matrix element:
\begin{align}
    b_y &= -\eta \frac{\phi}{\Omega} \left[ \sin\left(\frac{\phi}{2}\right) \sin(x_1 + x_3)  \right.\nonumber\\ 
    &\quad \left. +2\frac{\Delta}{\Omega}\cos\left(\frac{\phi}{2}\right) \sin x_1 \sin x_3 \right] = \eta\frac{\sin\theta}{\sqrt{2}}
    \label{eq:bsb_by_arb}
\end{align}

\section{Open-System Simulation}\label{ap:open_system}
Given the strong, non-adiabatic driving and the breakdown of the rotating-wave approximation (RWA) inherent to bang-bang control, standard open-system methodologies, such as the conventional Lindblad master equation, become invalid. The standard Lindblad formalism relies heavily on the secular approximation (or rotating wave approximation), which assumes weak drives and well-resolved transition frequencies. To accurately model dissipation under strong non-adiabatic driving, we instead employ the Universal Lindblad Equation (ULE) \cite{nathan2020universal}. The ULE framework is highly advantageous in this regime: it does not rely on the secular approximation, yet it preserves the complete positivity and trace preservation (CPTP) of the density matrix. Under the ULE, the evolution of the system density matrix $\rho(t)$ is governed by:
\begin{equation}
    \partial_{t}\rho(t) = -i[H_{\mathcal{S}}(t)+\Lambda(t), \rho(t)] + \sum_{\lambda=1}^{N}\mathcal{D}_{\lambda}[\rho(t),t]
\end{equation}
where $H_{\mathcal{S}}(t)$ is the system Hamiltonian, $\Lambda(t)$ is a Hermitian Lamb shift correction, and the dissipator takes the standard form:
\begin{equation}
    \mathcal{D}_{\lambda}[\rho,t] \equiv L_{\lambda}(t)\rho L_{\lambda}^{\dagger}(t) - \frac{1}{2}\{L_{\lambda}^{\dagger}(t)L_{\lambda}(t), \rho\}
\end{equation}

In the ULE framework, evaluating time-dependent jump operators in the instantaneous eigenbasis assumes that the Hamiltonian evolves slowly compared to the bath correlation time, $\tau$ (Markovian approximation). For an Ohmic environment at $T \sim 15$ mK, this correlation time is $\tau \sim 0.5$ ns. However, our bang-bang control pulses rely on sub-nanosecond smoothing ramps, explicitly violating this adiabatic condition. Because the drive timescale matches the bath correlation time, the environment cannot dynamically thermalize to the rapidly shifting energy gaps during the discrete transitions. To accurately reflect this non-adiabatic regime and capture the continuous baseline decoherence, we approximate the bath interaction as static, evaluating the jump operators at the undriven configuration (the fluxonium sweet spot). Under this approximation, the jump operators are time-independent and take the form:
\begin{equation}
    L = \sum_{mn} \sqrt{2\pi\gamma J(E_{n}-E_{m})} X_{mn} |m\rangle\langle n|
\end{equation}
where $\hat{X}$ is the system operator coupling to the bath, $J(\omega)$ is the bath spectral density, and $\gamma$ is the coupling strength. 

We restrict our microscopic Markovian noise models to the two dominant decoherence channels. The first is dielectric loss, which drives energy relaxation and couples to the bath via the charge operator $\hat{n}$. This is modeled as an Ohmic bath:
\begin{equation}
    J_{\text{ohmic}}(\omega) = \frac{\omega \exp\left(-\frac{1}{2}\left(\frac{\omega}{\Lambda_c}\right)^2\right)}{1 - \exp\left(-\frac{\omega}{T_{env}}\right)}
\end{equation}
The second channel is pure dephasing. While first-order flux noise vanishes at the sweet spot due to parity symmetry, second-order flux variations remain. This is modeled phenomenologically via a pure dephasing jump operator, restricted to white noise ($J_{\text{white}} = 2\pi$):
\begin{equation}
    X_{\text{dephasing}} = \sum_k \left( \frac{\partial^2 \omega_k}{\partial \delta_e^2}\right) |k\rangle\langle k|
\end{equation}
The coupling strengths $\gamma$ for both channels were fitted to reproduce the experimental coherence times $T_1$ and $T_{2e}$ ($\sim 200-300\;\mu\text{s}$) reported in the literature for the respective fluxonium regimes.

To efficiently simulate the ULE, we map the system into Liouville space by vectorizing the density matrix ($N\times N \rightarrow N^2 \times 1$), such that $\partial_t\rho_{vec}(t)=\mathcal{L}_{vec}(t)\rho_{vec}(t)$. The corresponding superoperator takes the form:
\begin{align}
    &\quad \quad\quad\mathcal{L}_{vec}(t) = -i(\mathds{1} \otimes H(t) - H(t)^T \otimes \mathds{1}) \nonumber\\
    &+ \sum_\lambda \left( L_\lambda^* \otimes L_\lambda - \frac{1}{2} \left( \mathds{1} \otimes L_\lambda^\dagger L_\lambda + (L_\lambda^\dagger L_\lambda)^T \otimes \mathds{1} \right) \right)
\end{align}
where $\otimes$ denotes the Kronecker product and the Lamb shift is absorbed into $H(t)$. This vectorization enables the same numerical caching strategy utilized for the closed-system dynamics. By pre-diagonalizing the Liouvillian for discrete pulse segments ($\mathcal{L}_{vec} = V D V^{-1}$), we generate an exact open-system propagator $E = V e^{D \tau}V^{-1}$. The propagators of the cosine ramps are calculated by integrating the master equation, similarly to the closed-system case. This $d^2\times d^2$ superoperator maps $\text{vec}(\rho(0)) \rightarrow \text{vec}(\rho(t_g))$. We truncate this full propagator to a $4\times 4$ operator $E$ corresponding to the logical $\{|0\rangle, |1\rangle\}$ subspace, with dimension $d_1$ and projector $\mathds{1}_1$. The average gate fidelity is then evaluated as \cite{wood2018quantification}:
\begin{equation}
    F(E) = \frac{\frac{1}{d_1} \text{tr}[(\mathds{1}_1 \otimes \mathds{1}_1)E] + 1 - L_1}{d_1 + 1}
\end{equation}
where $d_1 = 2$ and the leakage rate $L_1$ is defined via the subspace retention:
\begin{equation}
    L_1 = 1 - {\rm tr}\left( \mathds{1}_1 E\left(\frac{\mathds{1}_1}{d_1}\right) \right) =1 - \frac{1}{2}\sum_{k,j\in\{0,1\}} P(k\rightarrow j)
\end{equation}
with $P(k\rightarrow j)$ representing the probability of the initial state $|k\rangle\langle k|$ transitioning to $|j\rangle\langle j|$.

Flux-driven systems are also subject to non-Markovian $1/f$ flux noise. This slow, low-frequency oscillation is particularly detrimental to bang-bang protocols: while the qubit is highly protected at the sweet spot, the flux biases applied during the drive expose the system to first-order flux noise. Because the $1/f$ fluctuations are orders of magnitude slower than the nanosecond gate times, the noise acts as a static DC offset ($\delta\phi_{\text{DC}}$) during any single gate execution. The effective dynamics are thus described by the ensemble-averaged propagator:
\begin{equation}
    \overline{E} = \int_{-\infty}^{\infty} E(\delta\phi_{\text{DC}}) \frac{1}{\sqrt{2\pi\sigma^2}} \exp\left(-\frac{\delta\phi_{\text{DC}}^2}{2\sigma^2}\right) d(\delta\phi_{\text{DC}})
\end{equation}
where the standard deviation is set to $\sigma = 5.21 \times 10^{-6}\; \Phi_0$ \cite{zhang2021universal}. Because standard Monte Carlo integration of this ensemble converges too slowly, we employ Gauss-Hermite quadrature, which provides a more efficient approximation using discrete weighted nodes:
\begin{equation}
    \overline{E} \approx \sum_{i=1}^{N_{nodes}} \frac{w_i}{\sqrt{\pi}} E(\sqrt{2}\sigma x_i)
\end{equation}
Remarkably, because the execution times of these time-optimal sequences are exceptionally brief, the errors stemming from the $1/f$ flux exposure are heavily suppressed, even for non-symmetric protocols. We determine that the dominant error in the fastest gates stems instead from fundamental dielectric loss.

\section{Fair gate time comparisons}\label{sec:fair_comparisons}

 To accurately benchmark the speed of lab-frame gates against resonant microwave control, we establish a unified metric. Resonant schemes benefit from instantaneous virtual-$Z$ gates, implemented by shifting the carrier signal phase in software. Conversely, lab-frame control lacks a rotating-frame phase degree of freedom. To synthesize arbitrary rotations, $Z$ gates must be implemented via finite idling intervals. To quantify the performance tradeoff of these idling intervals, we analyze the aggregate gate performance over the standard single-qubit Clifford group.

We define the average Clifford gate time $\overline{t_g}$ as
\begin{equation}
    \overline{t_g} = \frac{1}{24} \sum_{n=1}^{24} t_g({\cal C}_n),
\end{equation}
where $t_g({\cal C}_n)$ is the total duration of the $n$-th Clifford gate. The $24$ elements of the single-qubit Clifford group can be partitioned into three decomposition subsets:
\begin{itemize}
    \item $Z$ gates: $\{Z(\varphi_k)\}$, with rotations $\varphi_k=k\frac{\pi}{2}$, for $k=0,1,2,3$.
    \item $\pi/2$ gates: $\{ Z(\varphi_k)\frac{X}{2}Z(\varphi_j)\}$, for $k,j \in \{0,1,2,3\}$.
    \item $\pi$ gates: $X$, $\frac{X}{2}Z\left(\frac{\pi}{2}\right)$, $Z\left(\frac{\pi}{2}\right)\frac{X}{2}$, and $Z\left(\frac{\pi}{2}\right)\frac{X}{2}Z\left(-\frac{\pi}{2}\right)$.
\end{itemize}
where the transverse axis $X/2$ can be equivalently substituted with $Y/2$, an operation that simply permutes the Clifford elements. The necessity of these $Z$ rotations intrinsically penalizes the average gate time of lab-frame protocols.

We evaluate $\overline{t_g}$ for three distinct control paradigms:

\subsection{Standard Resonant Control}

Using a native gate set equipped with instantaneous virtual-$Z$ gates and a $\pi/2$ gate of duration $t^{\text{R}}_{\pi/2}$, the subset decomposition yields $4$ gates with zero duration, $16$ gates requiring $t^{\text{R}}_{\pi/2}$, and $4$ gates requiring $2t^{\text{R}}_{\pi/2}$. This results in an average gate time of
\begin{equation}
    \overline{t_g^R} = t^{\text{R}}_{\pi/2}.
\end{equation}

\subsection{Commensurate pulses}
Commensurate driving schemes optimize native resonant \(X/2\) gates with durations matched to multiples of the counter-rotating period ($t_g = k \tau_L/2$), as detailed in \cite{rower2024suppressing}. \(Y/2\) gates are implemented by shifting the carrier phase by $\pi/2$ and appending $\tau_L/4$ idling periods, maintaining alignment with the counter-rotating lattice to ensure error suppression. The native gate set is restricted to $\mathcal{G}_C = \{I, \pm X/2, \pm Y/2\}$, where $t_g(\pm X/2)=t_{\pi/2}^C$. Commensurate idling intervals are utilized to synthesize the $Y$ gates: executing a $Y/2$ gate requires an initial $\tau_L/4$ idle to shift the drive oscillation phase by $\pi/2$, and a subsequent $\tau_L/4$ idle to return to the correct $X$ lattice coordinate, yielding $t_g(\pm Y/2)=t_{\pi/2}^C + \tau_L/2$.

The Clifford group is generated in this case via the composition
\begin{equation}
    {\cal C} = \left(\frac{X}{2}\right)^j \left(\frac{Y}{2}\right)^k \left(\frac{X}{2}\right)^l
\end{equation}
where $j, k \in \{0, 1, 2, 3\}$. If $k \in \{0, 2\}$, then $l = 0$; if $k \in \{1, 3\}$, then $l \in \{0, 1\}$. Weighting the required pulses across the $24$ Clifford elements yields an average duration composed of $52/24 \approx 2.167$ base gates and an accumulated lattice alignment penalty:
\begin{equation}
    \overline{t_g^C} \approx 2.167 \times t_{\pi/2}^C + 0.583 \times \tau_L.
\end{equation}

\subsection{Lab-frame control}

Our protocol employs the native generator set $\mathcal{G}_B = \{\pm Y/2, Z(\varphi_k)\}$. We optimize the compilation to minimize physical idling times by leveraging the commutation relation $Y/2 = Z(\pi)(-Y/2)Z(\pi)$. By absorbing $Z(\pi)$ rotations into the sign of the transverse drive, we selectively reduce the total accumulated idling phases required for the standard decomposition. Evaluating the gate sequences with this optimization yields an average gate time of
\begin{equation}
    \overline{t_g^B} = t^B_{\pi/2} + 0.4375 \times \tau_L.
\end{equation}

Ultimately, the evaluated average gate times across the benchmarked protocols are consolidated as follows:
\begin{itemize}
    \item Standard resonant: $\overline{t_g^R} = t^R_{\pi/2}$
    \item Commensurate resonant: $\overline{t_g^C} = 2.17 \times t_{\pi/2}^C + 0.58 \times \tau_L$
    \item Lab-frame control: $\overline{t_g^B} = t^B_{\pi/2} + 0.44 \times \tau_L$
\end{itemize}

\bibliography{bib/main.bib}
\end{document}